\documentclass{article} 
\usepackage{iclr2025_conference,times}

\usepackage{amsmath,amsfonts,bm}

\def\eqref#1{equation~\ref{#1}}

\def\1{\bm{1}}

\DeclareMathAlphabet{\mathsfit}{\encodingdefault}{\sfdefault}{m}{sl}
\SetMathAlphabet{\mathsfit}{bold}{\encodingdefault}{\sfdefault}{bx}{n}

\usepackage{hyperref}
\usepackage{url}

\usepackage{booktabs}       
\usepackage{pifont}
\usepackage{colortbl}
\usepackage{xcolor}
\usepackage{multirow}

\newcommand{\cmark}{\textcolor{green}{\ding{51}}} 
\newcommand{\xmark}{\textcolor{red}{\ding{55}}}   

\usepackage[disable]{todonotes}

\usepackage{subcaption}
\usepackage{makecell}
\usepackage{listings}
\usepackage{xcolor}
\usepackage{wrapfig}
\usepackage[normalem]{ulem}
\usepackage{hyperref}
\usepackage{cleveref}

\definecolor{codebg}{rgb}{0.95,0.95,0.95}
\definecolor{keyword}{rgb}{0.13,0.13,1}
\definecolor{comment}{rgb}{0,0.5,0}
\definecolor{string}{rgb}{0.56,0,0}

\newcommand{\GitChameleon}{\textcolor{violet}{\textbf{GitChameleon}}}

\title{\sout{\texttt{\textcolor{red}{df1.append(df2)}}} \texttt{\textcolor{teal}{pd.concat([df1,df2])}} GitChameleon: Unmasking the Version-Switching Capabilities of Code Generation Models}

\author{Nizar Islah, Justine Gehring \& Diganta Misra\textsuperscript{$\dagger$} \thanks{equal contribution} \hspace{26.5mm} Eilif Muller, Irina Rish\\
Mila - Quebec AI Institute  \hspace{58mm} Mila - Quebec AI Institute\\
MPI-IS Tübingen, ELLIS Tübingen\textsuperscript{$\dagger$}
\hspace{47mm} Université de Montréal\\
\texttt{\{firstname.lastname\}@mila.quebec} \\
\AND
Terry Yue Zhuo \hspace{80mm} Massimo Caccia\\
Monash University \hspace{72mm} ServiceNow Research \\
Sea AI Lab \\
}

\iclrfinalcopy 
\begin{document}

\maketitle

\begin{abstract}

  The rapid evolution of software libraries presents a significant challenge for code generation models, which must adapt to frequent version updates while maintaining compatibility with previous versions. Existing code completion benchmarks often overlook this dynamic aspect, and the one that does consider it relies on static code prediction tasks without execution-based evaluation, offering a limited perspective on a model's practical usability. To address this gap, we introduce \textbf{\GitChameleon{}}, a novel, manually curated dataset comprising 116 Python code completion problems, each conditioned on specific library versions and accompanied by executable unit tests. \GitChameleon{} is designed to rigorously assess the ability of modern large language models (LLMs) to generate version-specific code that is not only syntactically correct but also functionally accurate upon execution. Our comprehensive evaluations reveal that state-of-the-art LLMs struggle with this task; for instance, \textbf{GPT-4o} achieves a pass@10 of only 39.9\% (43.7\% when provided with error feedback), highlighting the complexity of the problem and the limitations of current models. By providing an execution-based benchmark that emphasizes the dynamic nature of code libraries, \GitChameleon{} serves as a critical tool to advance the development of more adaptable and reliable code generation models. For facilitation for further exploration of version-conditioned code generation, we make our code repository publicly accessible at \url{https://github.com/NizarIslah/GitChameleon}.

\end{abstract}

\section{Introduction}



\begin{figure}[h!]
    \centering
    \begin{minipage}{0.65\linewidth}
        \centering
        \includegraphics[width=\linewidth]{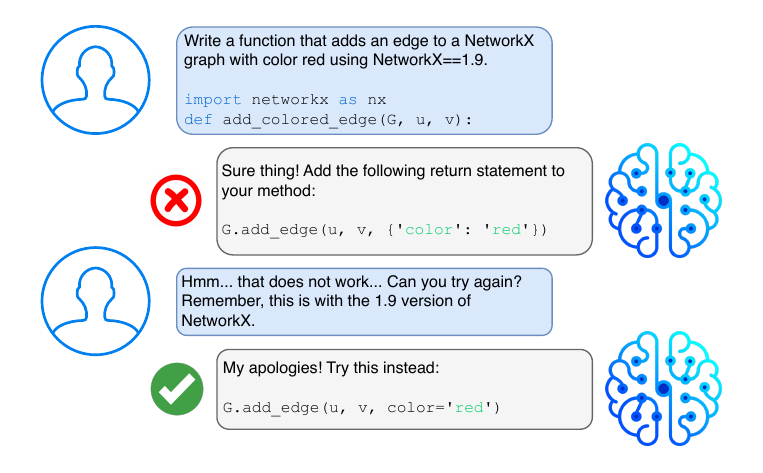}
        \label{fig:motivating_example}
    \end{minipage}%
    \hfill
    \begin{minipage}{0.33\linewidth}
        \centering
        \includegraphics[width=\linewidth]{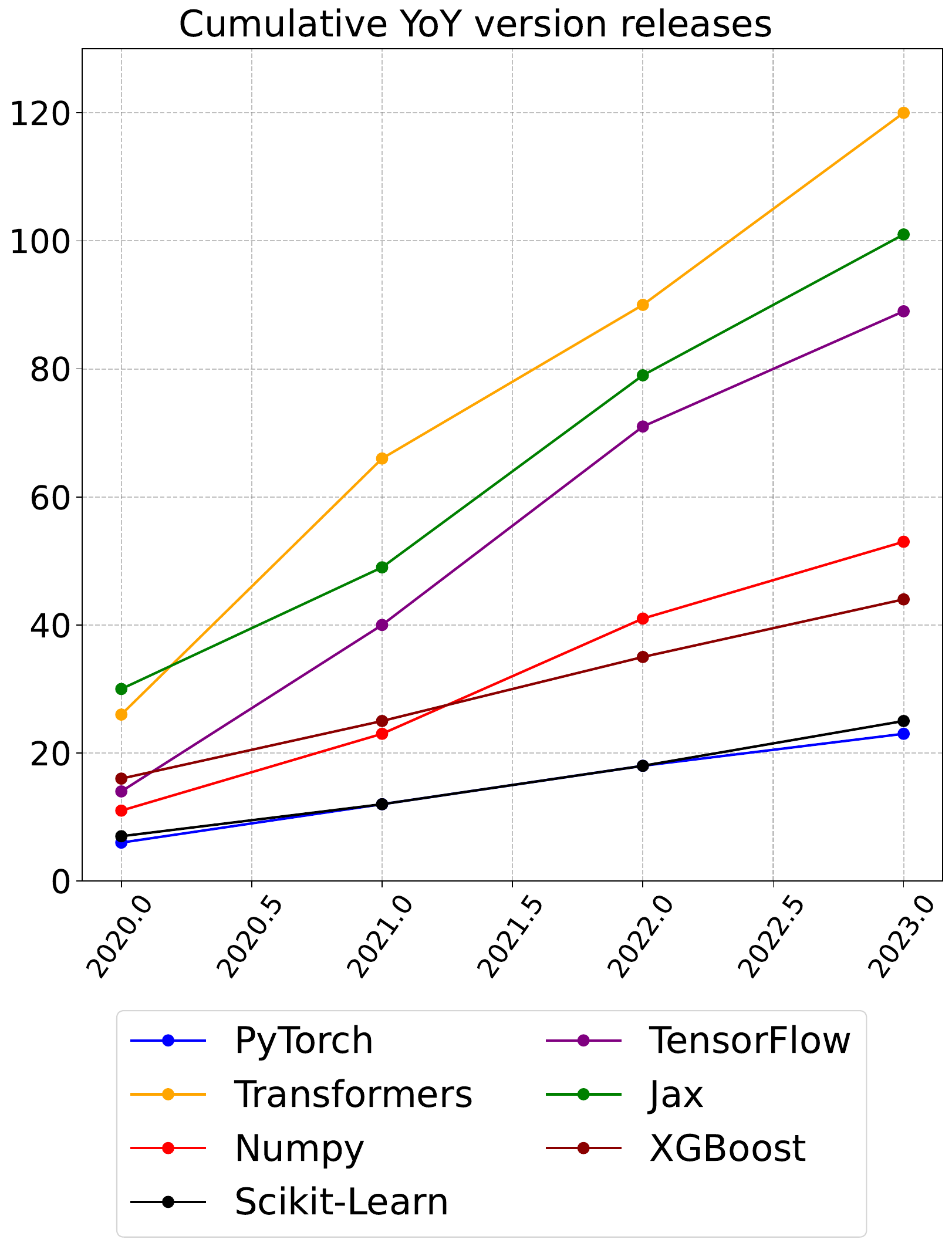}
        \label{fig:yoy_data}
    \end{minipage}
    \caption{\textbf{Left:} Modern LLMs often struggle with generating version-accurate code, highlighting the need for benchmarks that specifically assess their ability to handle versioning. \textbf{Right:} Cumulative year-over-year (YoY) version releases of popular Python-based machine learning libraries show a consistent upward trend, reflecting the rapid pace of development and version updates of code libraries and packages. 
    }
    \label{fig:fig1}
\end{figure}

Large Language Models (LLMs) have become highly popular in code completion, to the extent that they are now deployed as virtual coding assistants within popular code editors\footnote{\url{https://github.com/features/copilot}}, enhancing the overall coding workflow. 
Code, being a dynamic and constantly evolving environment, necessitates a continuous process of adaptation to stay in sync with the rapidly shifting paradigms, frameworks, and methodologies within the software development domain. The inherent variability in coding styles, the emergence of new programming languages, and the continuous evolution of libraries and packages underscore the imperative for an active approach in updating code generation models. 

In response to the needs of practical coding environments, several large language models (LLMs) have been introduced, including StarCoder \citep{li2023starcoder}, DeepSeek-Coder \citep{deepseek-coder}, CodeLlama \citep{roziere_code_2023}, among others. Despite these advancements, existing LLMs often struggle to keep pace with the rapid changes in codebases, particularly when tasked with generating version-specific code that is both syntactically and functionally accurate. This issue is especially critical, as developers increasingly depend on AI-assisted coding tools to boost productivity and maintain code quality. A recent Stack Overflow survey revealed that 70\% of the participants are using or planning to integrate AI coding tools, 33\% citing increased productivity as the primary motivation to incorporate these tools into their workflows\footnote{\url{https://stackoverflow.co/labs/developer-sentiment-ai-ml/}}.

Given the rapid development and release cycles of popular libraries, as shown in Figure \ref{fig:fig1} (right), the need for code generation models to continually adapt to changing API's is more pressing than ever. For example, prominent machine learning and deep learning libraries like \texttt{PyTorch} \citep{paszke2019pytorch}, \texttt{NumPy} \citep{harris2020array}, and \texttt{Scikit-Learn} \citep{sklearn_api} undergo frequent updates, which is reflected in a consistent upward trend in user downloads and version releases. This dynamic nature of code requires models that can adapt and generate code that adheres to the latest versions and practices, a need that current models often fail to meet comprehensively. 
In addition, certain hardware is restricted to compatibility with specific versions of commonly used packages, which adds an additional layer of complexity beyond merely updating the knowledge base of a code LLM to the latest library versions.

In response to these challenges, our work introduces a novel benchmark designed to assess the ability of LLMs to generate version-specific code. We propose \GitChameleon{}, a benchmark that evaluates state-of-the-art code models by requiring them to produce executable code based on version-specific prompts. Then, this code is executed to verify its correctness against expected outputs. By highlighting the limitations of current models in generating accurate version-specific code, \GitChameleon{} provides a structured approach to enhance these models and ensure their practical utility in real-world coding environments.

In summary, our contributions are highlighted as follows: 1) we introduce a novel code completion benchmark \GitChameleon{} consisting of 116 Python-based version conditioning problems, including human written unit tests; 2) we systematically analyze the version-specific performance of state-of-the-art code generation LLMs on API change type, version release year, and specific libraries. 3) we demonstrate the effectiveness of utilizing error log feedback as a way to improve version conditioning performance of code generation LLMs.

\section{\GitChameleon{} Benchmark}

We introduce \GitChameleon{}, a benchmark that comprises 116 Python-based version conditioning problems focused on popular code libraries. To evaluate LLM performance on GitChameleon, each problem is accompanied by handwritten assertion-based unit tests, enabling a thorough execution-based assessment of the outputs generated by the code LLMs. This structured approach enables a thorough understanding and categorization of LLM failures in common scenarios involving version-specific code generation problems. 
In the following sections, we detail the benchmark statistics, data collection methodology, and sample verification process.

\subsection{Statistics}

\GitChameleon{} consists of 116 python-based version conditioned problems based on 11 libraries: \texttt{PyTorch} \citep{paszke2019pytorch}, \texttt{Geopandas} \citep{kelsey_jordahl_2020_3946761}, \texttt{NLTK} \citep{nltk}, \texttt{NetworkX} \citep{hagberg2008exploring}, \texttt{GeoPy}\footnote{\url{https://pypi.org/project/geopy/}}, \texttt{Gradio} \citep{abid2019gradiohasslefreesharingtesting}, \texttt{Scikit-Learn} \citep{sklearn_api}, \texttt{Matplotlib} \citep{Hunter:2007}, \texttt{PyCaret}\footnote{\url{https://pycaret.org/}}, \texttt{Pandas} \citep{reback2020pandas,mckinney-proc-scipy-2010} and \texttt{NumPy} \citep{harris2020array}. The samples were collected from version releases over a period from the year 2014 to 2023 and exclude legacy and yanked version releases. 

\begin{table}[t]
  \caption{Compared to other popular code generation benchmarks, including those evaluating version conditioning, \GitChameleon{} features library- and version-specific problems with unit tests based on real version changes, closely aligning with practical settings.}
  \centering
  \scalebox{0.72}{ 
    \begin{footnotesize} 
    \begin{tabular}{@{}lcccccc@{}}
    \toprule
    Dataset   & Problems & Data Source   & Library Specific & Version Specific & Execution based & Real\\ \midrule
    HumanEval \citep{chen2021codex} & 164      & Hand-Written & \xmark  & \xmark & \cmark & - \\
    MBPP \citep{austin2021program}      & 974      & Hand-Written & \xmark & \xmark & \cmark & -  \\
    MTPB \citep{Nijkamp2022ACP} & 115 & Hand-Written & \xmark & \xmark & \cmark & -  \\
    APPS \citep{hendrycks2021measuring}     & 10000    & Competitions & \xmark & \xmark & \cmark &  -  \\
    CodeContests \citep{Li_2022} & 117 & Competitions & \xmark & \xmark & \cmark & -  \\
    JulCe \citep{agashe2019juice}    & 1518049     & Notebooks    & \xmark & \xmark & \cmark &  - \\
    DSP \citep{chandel2022training}  & 1119     & Notebooks   & \cmark   & \xmark & \cmark & - \\
    CoNaLa \citep{yin2018learning}   & 2879     & StackOverflow & \cmark & \xmark & \xmark & -  \\
    DS-1000 \citep{lai2023ds}  & 1000     & StackOverflow & \cmark & \xmark & \cmark & -  \\
    BigCodeBench \citep{zhuo2024bigcodebench}  & 1140     & Expert-Curated & \cmark & \xmark & \cmark & -  \\
    \midrule 
    Versicode \citep{wu2024versicodeversioncontrollablecodegeneration} &  98692  & GitHub, StackOverflow & \cmark & \cmark & \xmark & \cmark \\
    CodeUpdateArena \citep{liu2024codeupdatearenabenchmarkingknowledgeediting} & 670 & LLM-Generated & \cmark & \cmark & \cmark & \xmark \\
    \citep{wang2024llmsusedeprecatedapis} & 28125 & API change logs & \cmark & \cmark & \xmark & \cmark \\
    \midrule
    \rowcolor{gray!15}
    \GitChameleon{} \textbf{(Ours)} & 116 & Handwritten and LLM-assisted & \cmark & \cmark & \cmark & \cmark\\
    \bottomrule
    \end{tabular}
    \label{table:datasets}
    \end{footnotesize}
  }
\end{table}

Using the \texttt{cl100k\_base} tokenizer, we analyzed the token counts of the GitChameleon samples. The problem statements average 20.4 tokens, and the starter code averages 47.4 tokens, leading to a combined average of 67.8 tokens per sample. Including the prompt template utilized for evaluating instruction-tuned LLMs, the total token count across all samples is 19,409 tokens.

\begin{figure}[h]
    \centering
    \begin{subfigure}[b]{0.32\linewidth}
        \centering
        \includegraphics[width=\linewidth]{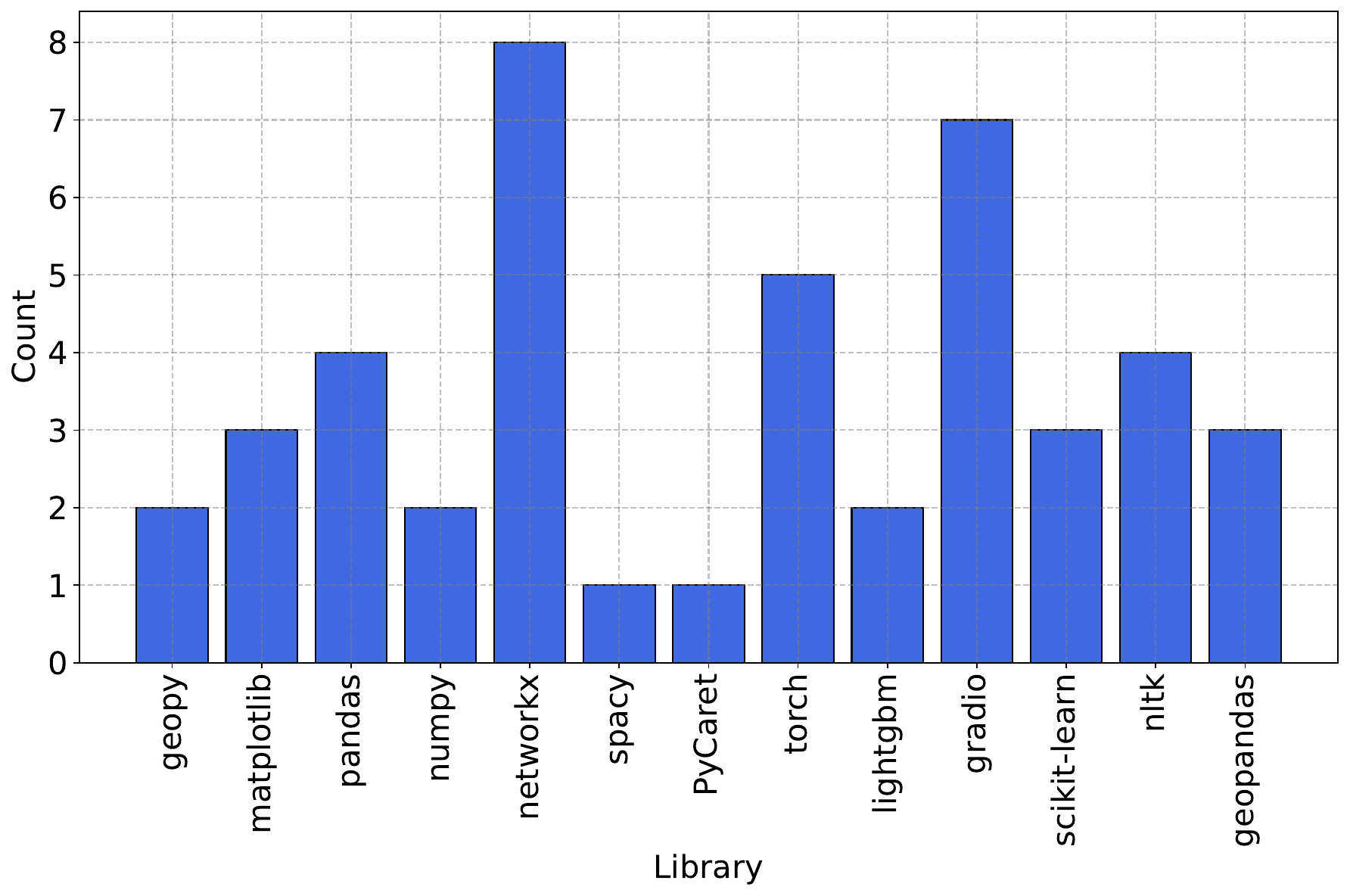}
        \caption{Number of unique versions per library in \GitChameleon{}.}
        \label{fig:version_count}
    \end{subfigure}
    \hfill
    \begin{subfigure}[b]{0.32\linewidth}
        \centering
        \includegraphics[width=\linewidth]{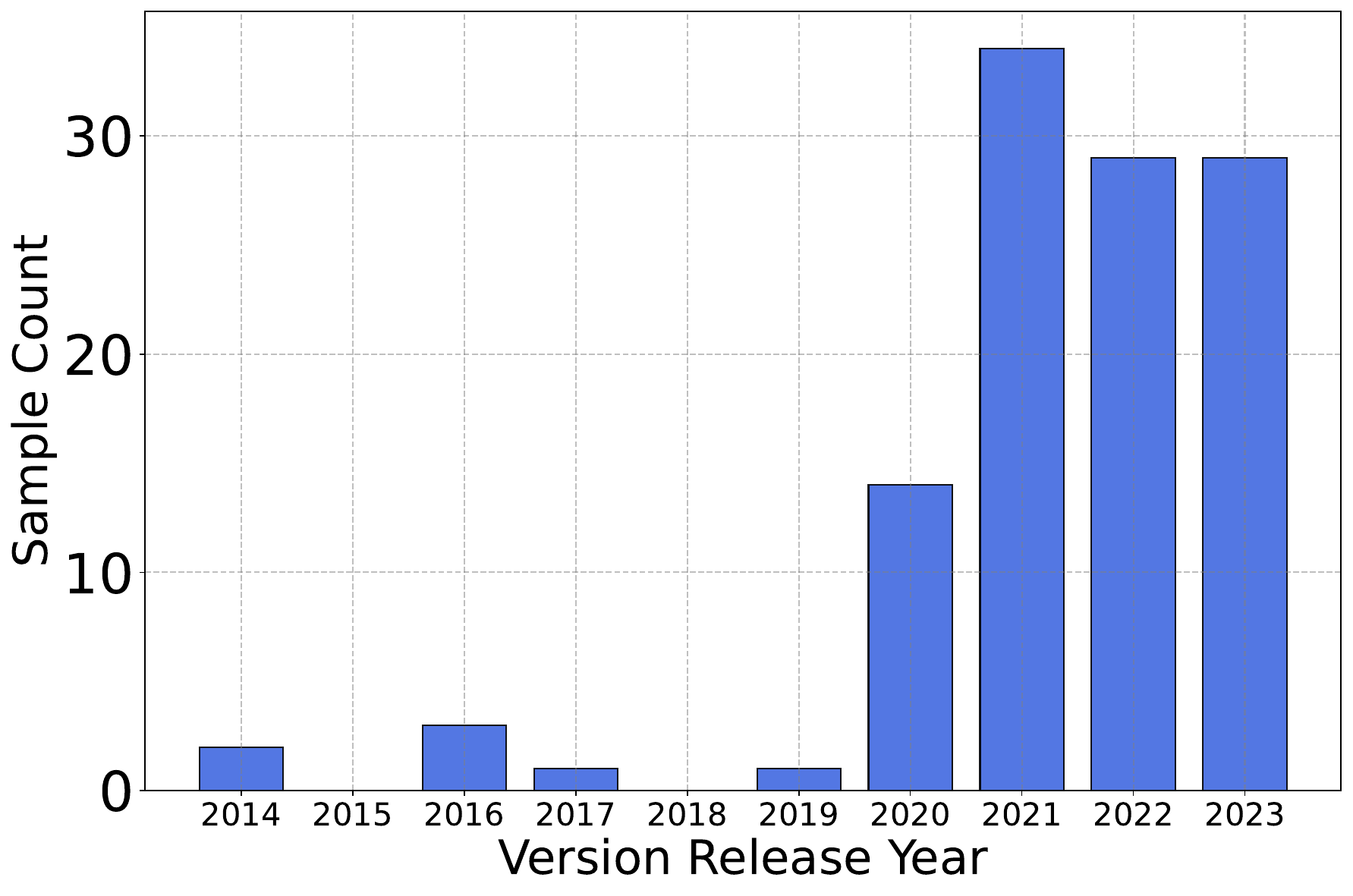}
        \caption{Number of samples by version release year.}
        \label{fig:version_year}
    \end{subfigure}
    \hfill
    \begin{subfigure}[b]{0.32\linewidth}
        \centering
        \includegraphics[width=\linewidth]{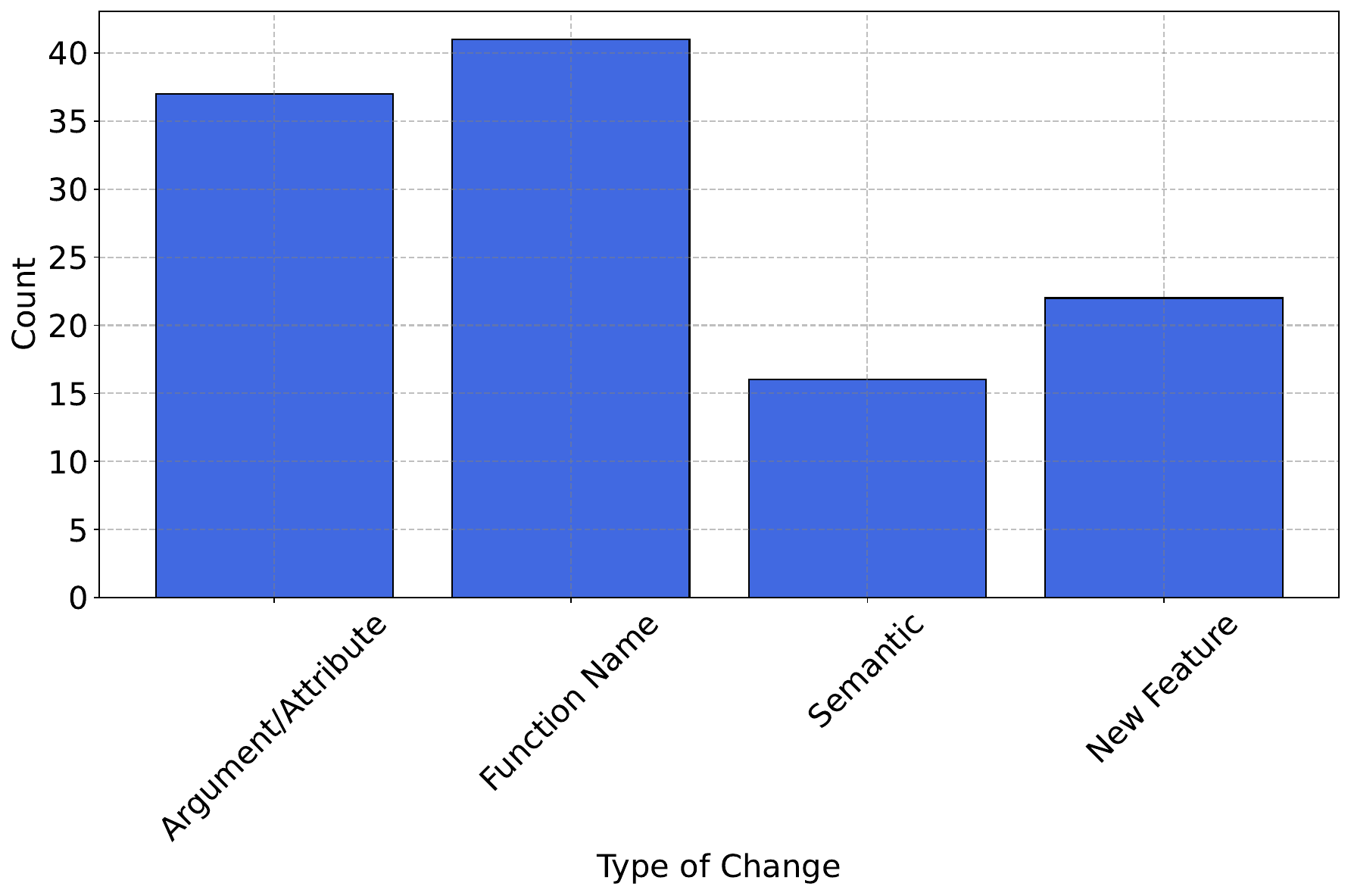}
        \caption{Number of samples by type of change.}
        \label{fig:change_type}
    \end{subfigure}
    \caption{Fine-grained statistics of the \GitChameleon{} benchmark. (a) The library with the most unique versions in the dataset is networkx with 8, whereas only 1 version of spacy and PyCaret are represented in \GitChameleon{}. (b) Most versions in the dataset were released between 2021-2023, with a few versions released in earlier years. (c) The most common type of changes between versions were function name changes and argument/attribute changes, while semantic/output changes were least common.}
\end{figure}

As demonstrated in Fig.~\ref{fig:version_year}, most of the samples in GitChameleon are from versions of libraries released in the years 2021, 2022 and 2023, with 2021 released version samples accounting for 35\% of the total sample count in the benchmark. Since some of the models evaluated on GitChameleon have disclosed their training data cutoff dates, we have ensured that most, if not all, samples fall within the training window of these models. 
This approach helps to ensure that the models during their training period are likely exposed to the versions on which the samples are based. Fig.~\ref{fig:version_count} shows that \texttt{NetworkX} \citep{hagberg2008exploring} and \texttt{Gradio} \citep{abid2019gradiohasslefreesharingtesting} have the most versions in our benchmark (8 and 7, respectively). Meanwhile, \texttt{PyTorch} \citep{paszke2019pytorch} and \texttt{NumPy} \citep{harris2020array} have the highest number of samples (18 and 15, respectively), together accounting for 34\% of the total sample count. In addition, we annotate each sample with the type of change that is classified into the following categories:
\begin{itemize}
    \item \textbf{Argument or Attribute change}: The API call to a function, method, or class has a change in arguments (e.g. name, order, new, deprecated argument) between versions.
    \item \textbf{Function Name change}: The name of the API call has changed between versions (e.g. \texttt{pandas.append} to \texttt{pandas.concat}).
    \item \textbf{Semantics or Function Behavior change}: The semantic / runtime behavior of the API call changed between versions (e.g. returning a different type).
    \item \textbf{New feature or additional dependency-based change}: A feature was introduced in a specific version; therefore, to execute the same functionality, a model using an older version should make use of an additional dependency (e.g. \texttt{torch.special} was introduced in \textsc{torch 1.10}, previously one could use \textsc{numpy} for the same).
\end{itemize}
Most samples in the GitChameleon benchmark fall under the Argument or Attribute and Function Name change category, as these are the most frequent and expected types of changes in mature and stable libraries.

\paragraph{Differentiating factor}
Several datasets examine LLM interactions with version-specific code, including Versicode \citep{wu2024versicodeversioncontrollablecodegeneration}, CodeUpdateArena \citep{liu2024codeupdatearenabenchmarkingknowledgeediting}, and the dataset by \citep{wang2024llmsusedeprecatedapis}. While these datasets are valuable, our dataset offers a unique and complementary perspective by focusing on the real-world scenario where developers are often constrained to specific library versions due to technical debt. CodeUpdateArena investigates model adaptation to synthetic API changes, we focus our evaluation on real API changes to assess how effectively an LLM can generate code for version-specific changes of library versions that they have been trained with. In contrast, Versicode and \cite{wang2024llmsusedeprecatedapis}’s datasets, while addressing library evolution, primarily rely on string matching for evaluation. Our approach diverges by incorporating executable tests, providing a more practical and rigorous assessment of code generation capabilities.

\subsection{Collection framework}

\begin{table*}[h]
\centering
\small
\renewcommand{\arraystretch}{1.1}
\setlength{\tabcolsep}{5pt} 
\begin{tabular}{p{6.4cm} p{6.4cm}} 
\textbf{Task Description} & \textbf{Expected Result} \\

    \begin{minipage}[t]{\linewidth} 
    \vspace{-1.5ex} 
        \begin{lstlisting}[language=Python, basicstyle=\ttfamily\scriptsize, frame=none, numbers=none]
    # Write a function that checks if all elements in an array are true.
    \end{lstlisting}
    \end{minipage} & 

    \begin{minipage}[t]{\linewidth} 
    \vspace{-1.5ex} 
    \begin{lstlisting}[language=Python, basicstyle=\ttfamily\scriptsize, frame=none, numbers=none]
    np.all(arr)
    \end{lstlisting}
    \vspace{1.5ex} 
    \end{minipage} \\

\textbf{Starter Code} & \textbf{Assertion Test} \\

    \begin{minipage}[t]{\linewidth} 
    \vspace{-1.5ex} 
    \begin{lstlisting}[language=Python, basicstyle=\ttfamily\scriptsize, frame=none, numbers=none]
    import numpy as np 
    def alltrue_fn(arr):
        return
    \end{lstlisting}
    \vspace{-1.5ex} 
    \end{minipage} & 

    \begin{minipage}[t]{\linewidth} 
    \vspace{-1.5ex} 
    \begin{lstlisting}[language=Python, basicstyle=\ttfamily\scriptsize, frame=none, numbers=none]
    arr = np.array([1, 1, 1, 1]) 
    result = alltrue_fn(arr) 
    assert result == np.all(arr)
    \end{lstlisting}
    \vspace{-1.5ex} 
    \end{minipage} \\ 

\end{tabular}
\caption{Example of a problem statement derived from a changelog entry from Numpy 1.25.0}
\label{tab:example}
\end{table*}

The examples were manually crafted by the authors, who divided the task among themselves. We compiled a list of popular Python libraries, focusing on those with which at least one author was familiar and that had detailed changelogs documenting changes between versions. For each library, we reviewed the changelogs to identify deprecated functions, argument changes, alterations in behavior, and newly introduced functions.

For each identified change, we create a concise problem statement, write the starter code, define the expected solution, and develop an assertion test. For instance, in Table \ref{tab:example}, we illustrate an example based on the changelog for version 1.25.0 of \texttt{NumPy} \citep{harris2020array}, a library for scientific computing in Python. This changelog notes that ``\texttt{np.alltrue} is deprecated. Use \texttt{np.all} instead." We used this change to craft a problem statement that tests the LLMs' ability to recognize and adapt to version-specific updates.

\paragraph{Unit test and evaluation framework verification}


To assess the correctness of the evaluation framework of GitChameleon, we needed to verify three key aspects: 
\begin{itemize}
    \item \textbf{Compilation}: Ensure that the starter code compiles successfully.
    \item \textbf{Assertion unit tests}: Confirm that the assertion tests function correctly.
    \item \textbf{Dependencies}: Verify that all necessary external dependencies are installed, excluding the ones being tested.
\end{itemize}

We used \texttt{venv} to create and manage virtual environments for testing. This process involved installing the appropriate library version and any additional dependencies. We then combined the starter code, expected result, and the assertion test into a single script, which was executed to verify all three criteria. We provide pseudocode for our verification process in \cref{app:verification_pseudocode}. 

\section{Empirical Study}

We evaluate state-of-the-art large language models (LLMs) using the \GitChameleon{} benchmark to assess their ability to generate version-specific, executable code. This study highlights how well current models adapt to dynamic library versions and produce functionally correct code that passes the provided unit tests.

\subsection{Experimental setup}

For each open-source LLM, we downloaded the corresponding Hugging Face (HF) weights and served the models using Text Generation Inference (TGI) for non-instruct models and VLLM~\citep{kwon2023efficient} for instruction tuned models. 
We used one NVIDIA 95GB H100 GPU for models with fewer than 70 billion parameters, two GPUs for models more than 70 billion parameters.

We configured the generation parameters with a \texttt{top\_p} value of 0.95, \texttt{top\_k} of 50, and a temperature of 0.3 for \textbf{Pass@1} and 0.8 for \textbf{Pass@10}.
The maximum number of new tokens generated was set to 256. 
Additionally, we enabled flash attention \citep{dao2022flashattentionfastmemoryefficientexact} for all models to enhance inference efficiency. 

\subsection{Evaluation Metrics}

To comprehensively evaluate the performance of code generation models using the \GitChameleon{} dataset, we employ a range of execution-based metrics. These metrics assess not only the correctness of the generated code, but also its efficiency and adaptability to different versions.

\textbf{\texttt{Pass@k}} measures the proportion of problems for which at least one of the top k generated solutions passes all assertion tests. This metric provides insight into the model's ability to generate functionally correct code. For each problem, we generate n code samples, and compute the pass at k metric by the corrected formula:
\begin{lstlisting}[language=Python]
    def corrected_pass_at_k(n, c, k=10):
    if n - c < k: return 1.0
    return 1.0 - np.prod(1.0 - k / np.arange(n - c + 1, n + 1))
\end{lstlisting}
For instruct models, we run the model's parsed output as standalone code, and for base models, the concatenation of the starting code and model's parsed output (completion).

\textbf{\texttt{Greedy}} refers to the standard greedy decoding method, where the most probable token from the next-token distribution is chosen deterministically. This is analogous to setting the temperature to 0.


\textbf{\texttt{Error Feedback}} adds the error log to the prompt (after executing the generated code from the model with the initial prompt). Then, the pass@k metric is recalculated based on the model's generated code using the prompt with error feedback. See \cref{app:example_err_feedback} for an example. 

\textbf{\texttt{doc-prompting}} is a method in which the documentation of the library function (specific to the version that the model is asked to use in the prompt) is given in the prompt. See Appendix A.1 for an example. 

\subsection{Main Results}

We report the performance of both base and instruct-tuned models on the \GitChameleon{} benchmark in Tables \ref{tab:base_model_performance} and \ref{tab:instruct_models_performance}, respectively. Our analysis reveals a positive correlation between model size and performance in version-specific code generation tasks. For base models, Spearman's rank correlation coefficients are 0.82 for Pass@1 and 0.69 for Pass@10 (both p-values $<$0.01), indicating that larger models generally perform better. Specifically, DeepSeek-Coder 33B achieved the highest Pass@1 score of 35.7\%, highlighting its proficiency in generating correct solutions on the first attempt, while CodeLlama 34B outperformed others at Pass@10 with a score of 42.8\%, demonstrating its ability to produce correct solutions given multiple attempts.

\begin{table*}[h!]
    \centering
    \resizebox{0.5\textwidth}{!}{
    \begin{tabular}{lcccc}
        \toprule
        \textbf{Model} & \textbf{Size} & \multicolumn{1}{c}{\textbf{Pass@1}} & \multicolumn{1}{c}{\textbf{Pass@10}} \\
        & & $$T=0.3$$ & $$T=0.8$$ \\
        \midrule
        \multirow{3}{*}{CodeLlama-Python \citep{rozière2024code} } & 7B  & 20.4$\pm$\scriptsize{\textcolor{gray}{1.6}} &   36.1$\pm$\scriptsize{\textcolor{gray}{5.5}} \\
        & 13B & 25.8$\pm$\scriptsize{\textcolor{gray}{1.0}} &  36.4$\pm$\scriptsize{\textcolor{gray}{2.0}} \\
        & 34B & 30.6$\pm$\scriptsize{\textcolor{gray}{1.4}} &  \textbf{42.8$\pm$\scriptsize{\textcolor{gray}{1.4}}} \\
        \midrule
        \multirow{3}{*}{Starcoder2 \citep{lozhkov2024starcoder2stackv2}}  & 3B & 11.9$\pm$\scriptsize{\textcolor{gray}{1.9}} &   27.1$\pm$\scriptsize{\textcolor{gray}{1.9}} \\
        & 7B & 15.5$\pm$\scriptsize{\textcolor{gray}{1.1}} &   23.1$\pm$\scriptsize{\textcolor{gray}{2.6}} \\
        & 15B & 13.7$\pm$\scriptsize{\textcolor{gray}{1.7}} &  27.0$\pm$\scriptsize{\textcolor{gray}{3.4}} \\
        \midrule
        \multirow{2}{*}{Llama{-}3 \citep{dubey2024llama3herdmodels}} 
         & 8B & 22.3$\pm$\scriptsize{\textcolor{gray}{1.0}}  & 32.0$\pm$\scriptsize{\textcolor{gray}{2.1}} \\
        & 70B & 27.2$\pm$\scriptsize{\textcolor{gray}{3.0}} &   41.3$\pm$\scriptsize{\textcolor{gray}{2.5}} \\
        \midrule
        \multirow{2}{*}{Qwen2 \citep{yang2024qwen2technicalreport} } & 7B & 27.4$\pm$\scriptsize{\textcolor{gray}{1.2}}  & 37.7$\pm$\scriptsize{\textcolor{gray}{1.8}} \\
        & 72B & 33.2$\pm$\scriptsize{\textcolor{gray}{2.1}} &  39.7$\pm$\scriptsize{\textcolor{gray}{5.5}} \\
        \midrule
        \multirow{4}{*}{Starcoderbase \citep{li2023starcoder}}  & 1B & 13.3$\pm$\scriptsize{\textcolor{gray}{1.0}} &  20.3$\pm$\scriptsize{\textcolor{gray}{1.2}} \\
        & 3B & 15.5$\pm$\scriptsize{\textcolor{gray}{1.2}} &   26.5$\pm$\scriptsize{\textcolor{gray}{1.5}} \\
        & 7B & 20.0$\pm$\scriptsize{\textcolor{gray}{0.9}} & 31.3$\pm$\scriptsize{\textcolor{gray}{4.1}} \\
        & 15B & 16.9$\pm$\scriptsize{\textcolor{gray}{1.8}} & 30.8$\pm$\scriptsize{\textcolor{gray}{2.6}} \\
        \midrule
        Starcoder \citep{li2023starcoder} & 15B & 16.0$\pm$\scriptsize{\textcolor{gray}{1.2}} & 35.9$\pm$\scriptsize{\textcolor{gray}{1.9}} \\
        \midrule
        \multirow{3}{*}{Deepseek{-}coder \citep{deepseek-coder} } & 1.3B & 22.0$\pm$\scriptsize{\textcolor{gray}{2.5}} & 28.0$\pm$\scriptsize{\textcolor{gray}{1.9}}  \\
        & 6.7B & 31.0$\pm$\scriptsize{\textcolor{gray}{1.8}} & 36.1$\pm$\scriptsize{\textcolor{gray}{0.7}} \\
        & 33B & \textbf{35.7$\pm$\scriptsize{\textcolor{gray}{3.0}}} & 37.9$\pm$\scriptsize{\textcolor{gray}{4.9}} \\
        \bottomrule
    \end{tabular}}
    \caption{
    \textbf{Base Model Performance Metrics.} 
        Deepseek-coder-33B is the strongest model for Pass@1 (temperature 0.3), while CodeLlama-34B is the strongest model when we compute Pass@10 with an increased number of generations (20) sampled at temperature 0.8. We observe that there is a strong positive correlation between model size and performance, with Spearman's rank correlation coefficients of 0.82 for Pass@1 and 0.69 for Pass@10.
        }
    \label{tab:base_model_performance}
\end{table*}

Similarly, for instruct-tuned models, we observe Spearman's rank correlation coefficients of 0.45 for Pass@1 and 0.42 for Pass@10 (both with p-values under 1\%), confirming the positive correlation between model size and performance. Phi-3.5-MoE (16×3.8B) achieved the highest baseline Pass@1 score of 30.9\% (33.6\% greedy) and highest baseline Pass@10 (40.5\%) for OSS models. Overall, Gemini achieves the highest baseline pass@10 (41.4\%). Yi-1.5-Chat (34B) slightly outperformed GPT-4o and Gemini at Pass@10 with error feedback with a score of 44.1\% compared to 43.7\% and 43.8\%, respectively. Incorporating error feedback led to average improvements of 5.4\% in Pass@1 and 4.7\% in Pass@10 across instruct-tuned models. Additionally, Pass@10 with n=20 samples showed an average performance improvements of 10.6\% for base models and 14.8\% for instruct-tuned models over Pass@1 with n=5. These findings highlight that scaling model size, utilizing error feedback, and allowing multiple solution attempts are effective strategies for enhancing the ability of LLMs in handling version-specific code generation tasks.

\begin{table*}[h!]
    \centering
    \resizebox{\textwidth}{!}{
    \begin{tabular}{lccccccc}
        \toprule
        \textbf{Model} & \textbf{Size (Context) /} & \multicolumn{4}{c}{\textbf{Pass@1} ($T=0.3$)} & \multicolumn{2}{c}{\textbf{Pass@10} ($T=0.8$)} \\
        \cmidrule(lr){3-6} \cmidrule(lr){7-8}
        & \textbf{Version} & Baseline & + Error Feedback & Greedy\textsuperscript{$\dagger$} & Zero-shot CoT & Baseline & + Error Feedback \\
        \midrule
        \multirow{1}{*}{Deepseek-Coder-V2-Lite} \citep{deepseekai2024deepseekcoderv2breakingbarrierclosedsource} & 16B & 27.4$\pm$\scriptsize{\textcolor{gray}{1.0}} & 33.4$\pm$\scriptsize{\textcolor{gray}{0.5}} & 26.7 & 26.7 & 40.3$\pm$\scriptsize{\textcolor{gray}{0.7}} & 43.5$\pm$\scriptsize{\textcolor{gray}{0.9}} \\
        \midrule
        \multirow{1}{*}{Starcoder2{-}v0.1} \citep{lozhkov2024starcoder2stackv2} & 15B & 22.0$\pm$\scriptsize{\textcolor{gray}{0.9}} & 27.9$\pm$\scriptsize{\textcolor{gray}{1.6}} & 21.6 & 28.4 & 37.8$\pm$\scriptsize{\textcolor{gray}{0.3}} & 39.9$\pm$\scriptsize{\textcolor{gray}{0.3}} \\
        \midrule
        \multirow{3}{*}{CodeLlama \citep{rozière2024code}}  & 7B & 16.4$\pm$\scriptsize{\textcolor{gray}{1.1}} & 18.8$\pm$\scriptsize{\textcolor{gray}{1.7}} & 19.0 & 0.9 & 29.0$\pm$\scriptsize{\textcolor{gray}{1.5}} & 33.1$\pm$\scriptsize{\textcolor{gray}{1.5}} \\
        & 13B & 20.3$\pm$\scriptsize{\textcolor{gray}{0.6}} & 25.6$\pm$\scriptsize{\textcolor{gray}{1.3}} & 22.4 &0.0 &  35.7$\pm$\scriptsize{\textcolor{gray}{0.8}} & 41.2$\pm$\scriptsize{\textcolor{gray}{1.0}} \\
          & 34B & 17.4$\pm$\scriptsize{\textcolor{gray}{0.8}} & 23.3$\pm$\scriptsize{\textcolor{gray}{1.1}} & 18.1 & 3.4&34.7$\pm$\scriptsize{\textcolor{gray}{1.0}} & 39.9$\pm$\scriptsize{\textcolor{gray}{1.9}} \\
        \midrule
        \multirow{1}{*}{Llama{-}3.1 \citep{dubey2024llama3herdmodels}}  & 8B & 16.0$\pm$\scriptsize{\textcolor{gray}{0.4}} & 20.5$\pm$\scriptsize{\textcolor{gray}{0.7}} & 17.2 & 19.8 & 29.2$\pm$\scriptsize{\textcolor{gray}{0.5}} & 34.9$\pm$\scriptsize{\textcolor{gray}{0.6}} \\
        \midrule
        \multirow{2}{*}{Llama{-}3.2 \citep{dubey2024llama3herdmodels}}  & 1B & 9.7$\pm$\scriptsize{\textcolor{gray}{0.4}} & 13.0$\pm$\scriptsize{\textcolor{gray}{0.8}} & 9.5 & 3.4 & 16.2$\pm$\scriptsize{\textcolor{gray}{0.8}} & 21.1$\pm$\scriptsize{\textcolor{gray}{0.6}} \\
        & 3B & 10.6$\pm$\scriptsize{\textcolor{gray}{0.8}} & 14.5$\pm$\scriptsize{\textcolor{gray}{1.0}} & 12.1 & 11.2 & 20.2$\pm$\scriptsize{\textcolor{gray}{1.5}} & 23.9$\pm$\scriptsize{\textcolor{gray}{2.7}} \\
        \midrule
        \multirow{1}{*}{CodeQwen1.5-Chat \citep{bai2023qwentechnicalreport}} 
        & 7B & 21.4$\pm$\scriptsize{\textcolor{gray}{0.8}} & 25.9$\pm$\scriptsize{\textcolor{gray}{0.6}} & 19.0 & 27.6 & 38.6$\pm$\scriptsize{\textcolor{gray}{1.5}} & 42.4$\pm$\scriptsize{\textcolor{gray}{0.8}} \\
        \midrule
        \multirow{2}{*}{Qwen2 \citep{yang2024qwen2technicalreport}}  & 7B & 18.3$\pm$\scriptsize{\textcolor{gray}{0.7}} & 26.7$\pm$\scriptsize{\textcolor{gray}{1.5}} & 19.0 &19.0 & 38.0$\pm$\scriptsize{\textcolor{gray}{0.6}} & 43.3$\pm$\scriptsize{\textcolor{gray}{0.5}} \\
         & 72B & 25.0$\pm$\scriptsize{\textcolor{gray}{0.7}} & 30.1$\pm$\scriptsize{\textcolor{gray}{0.3}} & 23.3 &  24.1 & 38.2$\pm$\scriptsize{\textcolor{gray}{0.9}} & 40.7$\pm$\scriptsize{\textcolor{gray}{0.8}} \\
        \midrule
        \multirow{2}{*}{Qwen2.5-Coder \citep{hui2024qwen25codertechnicalreport}} 
        & 1.5B & 19.7$\pm$\scriptsize{\textcolor{gray}{0.8}} & 22.3$\pm$\scriptsize{\textcolor{gray}{1.2}} & 19.8 & 17.2 & 32.7$\pm$\scriptsize{\textcolor{gray}{1.0}} & 38.8$\pm$\scriptsize{\textcolor{gray}{0.7}} \\
        & 7B & 21.1$\pm$\scriptsize{\textcolor{gray}{0.5}} & 24.4$\pm$\scriptsize{\textcolor{gray}{0.4}} & 22.4 & 14.7 & 35.6$\pm$\scriptsize{\textcolor{gray}{1.3}} & 41.5$\pm$\scriptsize{\textcolor{gray}{0.7}} \\
        \midrule
        \multirow{1}{*}{Codestral-v0.1 (https://mistral.ai/news/codestral/)} & 22B & 25.0$\pm$\scriptsize{\textcolor{gray}{0.4}} & 31.3$\pm$\scriptsize{\textcolor{gray}{0.8}} & 24.1 &34.5 & 37.1$\pm$\scriptsize{\textcolor{gray}{0.8}} & 40.6$\pm$\scriptsize{\textcolor{gray}{0.9}} \\
          \midrule
        \multirow{2}{*}{Yi-Coder-Chat \citep{ai2024yiopenfoundationmodels}} & 1.5B & 18.7$\pm$\scriptsize{\textcolor{gray}{0.5}} & 22.0$\pm$\scriptsize{\textcolor{gray}{0.9}} & 18.1 &21.6 &  32.6$\pm$\scriptsize{\textcolor{gray}{1.7}} & 37.0$\pm$\scriptsize{\textcolor{gray}{1.7}} \\
        & 9B & 25.7$\pm$\scriptsize{\textcolor{gray}{0.5}} & 30.8$\pm$\scriptsize{\textcolor{gray}{0.6}} & 25.0 &22.4 & 38.6$\pm$\scriptsize{\textcolor{gray}{0.9}} & 41.5$\pm$\scriptsize{\textcolor{gray}{0.8}} \\
        \midrule
        \multirow{3}{*}{Yi-1.5-Chat \citep{ai2024yiopenfoundationmodels}} & 6B & 17.3$\pm$\scriptsize{\textcolor{gray}{1.0}} & 23.9$\pm$\scriptsize{\textcolor{gray}{1.7}} & 15.5 & 19.0 & 32.8$\pm$\scriptsize{\textcolor{gray}{1.0}} & 37.4$\pm$\scriptsize{\textcolor{gray}{1.5}} \\
        & 9B & 20.1$\pm$\scriptsize{\textcolor{gray}{0.3}} & 24.7$\pm$\scriptsize{\textcolor{gray}{1.0}} & 19.8 &21.6& 30.5$\pm$\scriptsize{\textcolor{gray}{0.6}} & 39.5$\pm$\scriptsize{\textcolor{gray}{1.1}} \\
        & 34B & 20.8$\pm$\scriptsize{\textcolor{gray}{0.5}} & 26.3$\pm$\scriptsize{\textcolor{gray}{1.0}} & 21.6 & 20.7 &35.3$\pm$\scriptsize{\textcolor{gray}{0.8}} & \textbf{44.1}$\pm$\scriptsize{\textcolor{gray}{0.7}} \\
        \midrule
        \multirow{1}{*}{codegemma \citep{codegemmateam2024codegemmaopencodemodels}} & 7B & 17.8$\pm$\scriptsize{\textcolor{gray}{0.8}} & 23.6$\pm$\scriptsize{\textcolor{gray}{0.2}} & 17.2 & 19.8 & 33.9$\pm$\scriptsize{\textcolor{gray}{0.6}} & 38.0$\pm$\scriptsize{\textcolor{gray}{0.5}} \\
        \midrule
        \multirow{1}{*}{stable-code \citep{pinnaparaju2024stablecodetechnicalreport}} & 3B & 14.6$\pm$\scriptsize{\textcolor{gray}{0.7}} & 16.3$\pm$\scriptsize{\textcolor{gray}{0.9}} & 14.7 & 17.2 & 23.9$\pm$\scriptsize{\textcolor{gray}{1.4}} & 25.9$\pm$\scriptsize{\textcolor{gray}{0.9}} \\
        \midrule 
        \multirow{5}{*}{granite-code \citep{mishra2024granitecodemodelsfamily}} & 3B (128k) & 23.9$\pm$\scriptsize{\textcolor{gray}{1.1}} & 26.1$\pm$\scriptsize{\textcolor{gray}{2.0}} & 22.4 & 29.3 & 33.6$\pm$\scriptsize{\textcolor{gray}{1.3}} & 36.5$\pm$\scriptsize{\textcolor{gray}{0.6}} \\
           & 8B (4k) & 24.7$\pm$\scriptsize{\textcolor{gray}{0.9}} & 29.2$\pm$\scriptsize{\textcolor{gray}{1.5}} & 24.1 &28.4 &  38.5$\pm$\scriptsize{\textcolor{gray}{0.4}} & 42.1$\pm$\scriptsize{\textcolor{gray}{1.0}} \\
            & 8B (128k) & 23.9$\pm$\scriptsize{\textcolor{gray}{0.7}} & 28.7$\pm$\scriptsize{\textcolor{gray}{1.1}} & 25.0 & 29.3 & 35.0$\pm$\scriptsize{\textcolor{gray}{0.5}} & 38.9$\pm$\scriptsize{\textcolor{gray}{0.6}} \\
             & 20B (8k) & 28.2$\pm$\scriptsize{\textcolor{gray}{0.6}} & 30.5$\pm$\scriptsize{\textcolor{gray}{0.8}} & 29.3 & 31.0 & 37.0$\pm$\scriptsize{\textcolor{gray}{0.7}} & 38.3$\pm$\scriptsize{\textcolor{gray}{0.8}} \\
              & 34B (8k) & 29.8$\pm$\scriptsize{\textcolor{gray}{0.9}} & 31.7$\pm$\scriptsize{\textcolor{gray}{1.2}} & 31.0 & 31.0 & 38.3$\pm$\scriptsize{\textcolor{gray}{0.1}} & 39.9$\pm$\scriptsize{\textcolor{gray}{1.3}} \\
        \midrule
        \multirow{1}{*}{Phi-3.5-mini} \citep{abdin2024phi3technicalreporthighly} & 3.8B & 24.0$\pm$\scriptsize{\textcolor{gray}{0.7}} & 29.0$\pm$\scriptsize{\textcolor{gray}{0.9}} & 25.9 & 19.8 & 35.1$\pm$\scriptsize{\textcolor{gray}{1.1}} & 39.2$\pm$\scriptsize{\textcolor{gray}{1.0}} \\
        \midrule 
        \multirow{1}{*}{Phi-3.5-MoE \citep{abdin2024phi3technicalreporthighly}} & 16x3.8B & \textbf{30.9$\pm$\scriptsize{\textcolor{gray}{0.8}}} & 34.9$\pm$\scriptsize{\textcolor{gray}{0.7}} & \textbf{33.6} &23.3 &  40.5$\pm$\scriptsize{\textcolor{gray}{0.5}} & 43.2$\pm$\scriptsize{\textcolor{gray}{0.1}} \\
        \midrule 
        \multirow{1}{*}{Nxcode-CQ-orpo} \citep{hong2024orpomonolithicpreferenceoptimization} & 7B & 20.8$\pm$\scriptsize{\textcolor{gray}{0.5}} & 26.3$\pm$\scriptsize{\textcolor{gray}{1.3}} & 19.8 & 27.6 & 38.7$\pm$\scriptsize{\textcolor{gray}{1.8}} & 43.2$\pm$\scriptsize{\textcolor{gray}{1.3}} \\
        \midrule 
        \midrule
         \multirow{2}{*}{GPT \citep{openai2024gpt4technicalreport}}     & 3.5  & 19.6$\pm$\scriptsize{\textcolor{gray}{1.0}} &27.2$\pm$\scriptsize{\textcolor{gray}{0.8}} & 19.8 & 20.7 & 33.3$\pm$\scriptsize{\textcolor{gray}{1.0}} & 37.8$\pm$\scriptsize{\textcolor{gray}{1.7}} \\
        & 4o (2024-05-13)  & 23.6$\pm$\scriptsize{\textcolor{gray}{2.7}} &34.1$\pm$\scriptsize{\textcolor{gray}{1.2}} & 25.0 & 27.6 & 39.9$\pm$\scriptsize{\textcolor{gray}{2.4}} & 43.7$\pm$\scriptsize{\textcolor{gray}{2.4}} \\
        \midrule
         \multirow{1}{*}{Gemini \citep{gemini15}}  & 1.5-pro-001 & 25.1$\pm$\scriptsize{\textcolor{gray}{0.2}} & \textbf{35.9}$\pm$\scriptsize{\textcolor{gray}{0.2}} & 32.8 & \textbf{37.1} & \textbf{41.4}$\pm$\scriptsize{\textcolor{gray}{0.5}} & 43.8$\pm$\scriptsize{\textcolor{gray}{0.6}} \\
    \bottomrule
    \end{tabular}}
    \caption{\textbf{Instruct Model performance metrics.} (Top) OSS models, (bottom) closed-sourced models. 
    We observe a 5.4\% and 4.7\% improvement with error feedback in Pass@1 and Pass@10, respectively.
    Additionally, there is a positive correlation between model size and performance, with Spearman's rank correlation coefficients of 0.45 for Pass@1 and 0.42 for Pass@10.
    }
    \label{tab:instruct_models_performance}. 
    \vspace{-5mm}
\end{table*}


\begin{figure}
    \centering
    \includegraphics[width=0.80\linewidth]{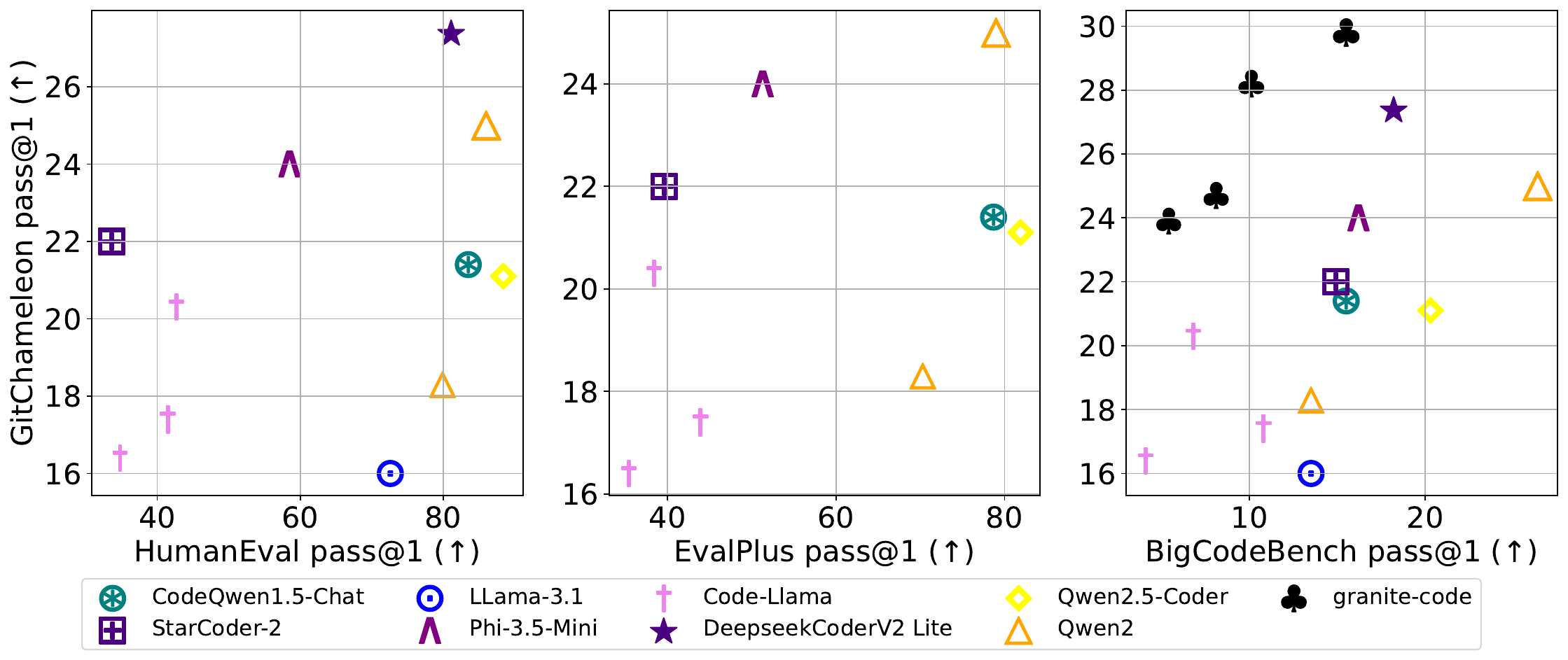}
    \caption{Correlation of \GitChameleon{} with the representative code benchmarks (HumanEval, EvalPlus, and BigCodeBench-Hard split). For each benchmark, the spearman correlation coefficient was 0.37, 0.50, and 0.36, respectively. }
    \label{fig:dataset-corr}
\end{figure}

To highlight the uniqueness of our proposed evaluation dataset and difficulty of the samples in \GitChameleon{}, we examine the performance correlation of models on \GitChameleon{} in Figure~\ref{fig:dataset-corr} with three standard code generation evaluation benchmarks: HumanEval~\citep{chen2021codex}, EvalPlus~\citep{liu2024your}, and the Hard split of BigCodeBench~\citep{zhuo2024bigcodebench}. Although performance on these benchmarks shows a positive correlation with that on \GitChameleon{}, the relationship is not strictly predictive, as models that perform well on one or more of these benchmarks do not necessarily achieve equivalent success on \GitChameleon{}.


\subsection{In-Depth Analysis of Findings}%
\label{ssec:ModelPerformanceMetrics}%

In this section, we delve deeper into the results obtained from our experiments, analyzing model performance across various dimensions, including model size, year of library release, the type of API changes encountered, and sample difficulty.

\begin{figure}[t]
    \centering
    \begin{minipage}{\linewidth}
        \centering
        \includegraphics[width=\linewidth]{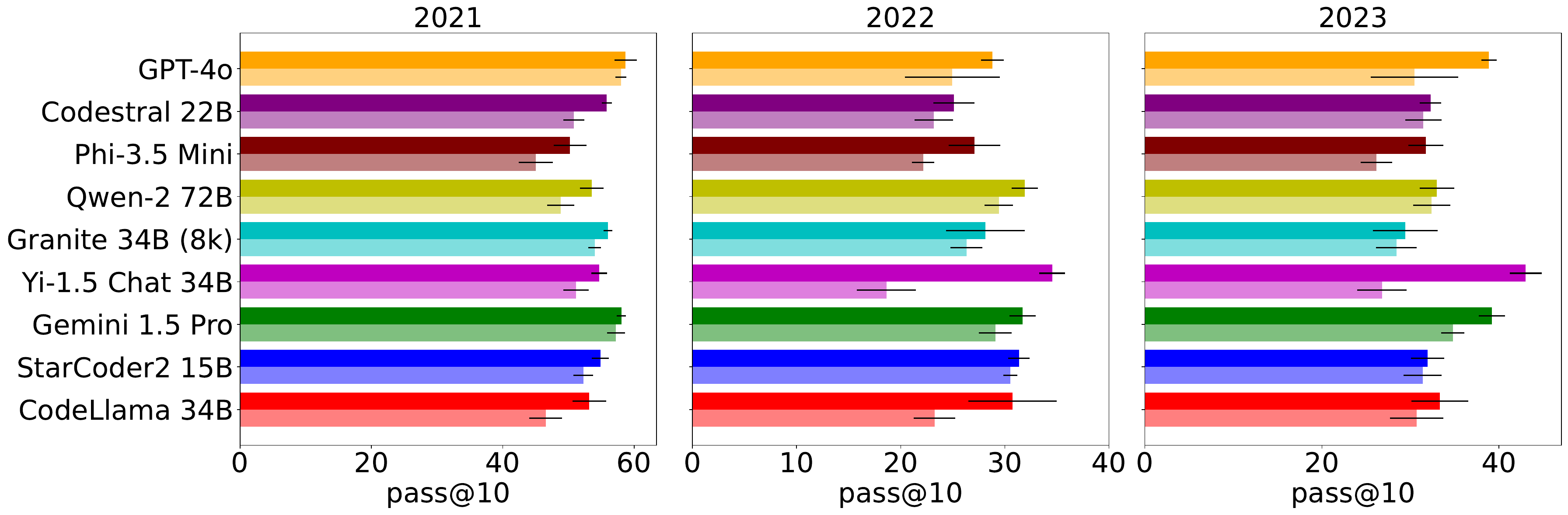}
        \label{fig:model_date_perf}
    \end{minipage}
    \vspace{5mm} 
    \begin{minipage}{\linewidth}
        \centering
        \includegraphics[width=\linewidth]{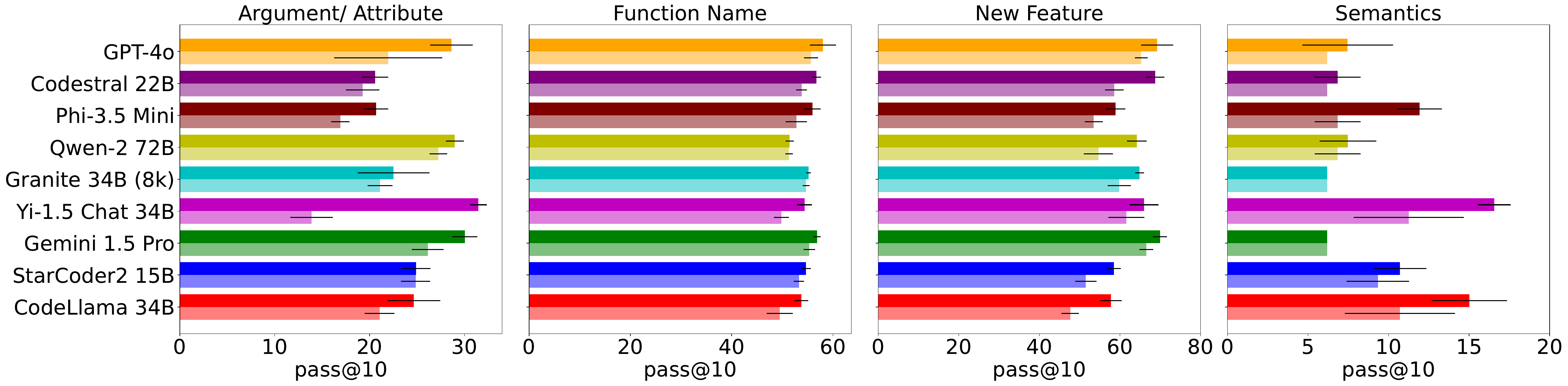}
        \label{fig:model_change_perf}
    \end{minipage}
    \caption{\textbf{Instruct-tuned model performance breakdown by version release year (top) and type of change (bottom)}: We analyze model performance in terms of \texttt{pass @ 10} for baseline and with error feedback generation across two dimensions: version release year and type of changes.
    Darker shaded bars represent values obtained via error feedback generation. Standard deviation is drawn as a black line, obtained from 5 random seeds. (Top) Many models perform poorly on 2022, and generally perform worse on more recent versions. (Bottom) All models perform very poorly at semantic changes, indicating an potential area for massive improvement. Most models perform well on function name changes and new feature (with the exception of Llama-3.2-3B, which is also the smallest model in this comparison).}
    \label{fig:combined_model_perf}
\end{figure}

\paragraph{Analysis of Performance by Release Date}


At the top of Figure \ref{fig:combined_model_perf}, we present the year-over-year performance of a subset of the instruction-finetuned models. The average performance of all models dropped significantly from 2019 (not shown) to 2023. This decline is likely due to the fact that the training datasets contain more data from earlier years, underscoring the need for code LLMs to better adapt to the evolving nature of code libraries and their versions. Interestingly, many models appear to improve with error feedback disproportionately across versions released in the years 2021-2023. For example, Qwen2-72B and CodeLlama-34B improve more with feedback in 2021-2022 compared to 2023, while GPT-4o and Yi-1.5 Chat 34B improve more with feedback in 2023. This raises a question about the extent to which models' training data influence the effectiveness of error feedback.


\paragraph{Analysis of Performance by Type of API Change}


At the bottom of Figure \ref{fig:combined_model_perf}, our analysis of model performance across different types of API changes in \GitChameleon{} revealed significant variations. The models struggled the most with \textbf{Semantics or Function Behavior} changes. 
\textbf{Argument and Attribute} changes were the second most challenging.
In contrast, models performed better on \textbf{Function Name} changes and \textbf{New Feature or additional dependency based} changes.

In general, larger models are more robust to function name, argument/attribute changes, and new feature changes. However, all models perform very poorly on semantic changes, regardless of the availability of error feedback. This indicates a weakness of SotA code generation models, and an area for further investigation. Furthermore, error feedback appears to have a more significant impact in argument/attribute changes compared to the other types of changes. This indicates that 1) the models may be using the error feedback to directly address failures in version-conditioned code generation, rather than non-version-specific errors, and 2) error feedback is most effective for version-specific argument/attribute changes.



\paragraph{Sample difficulty analysis}

\begin{figure}[t]
    \centering
    \begin{minipage}{0.49\linewidth}
        \centering
        \includegraphics[width=\linewidth]{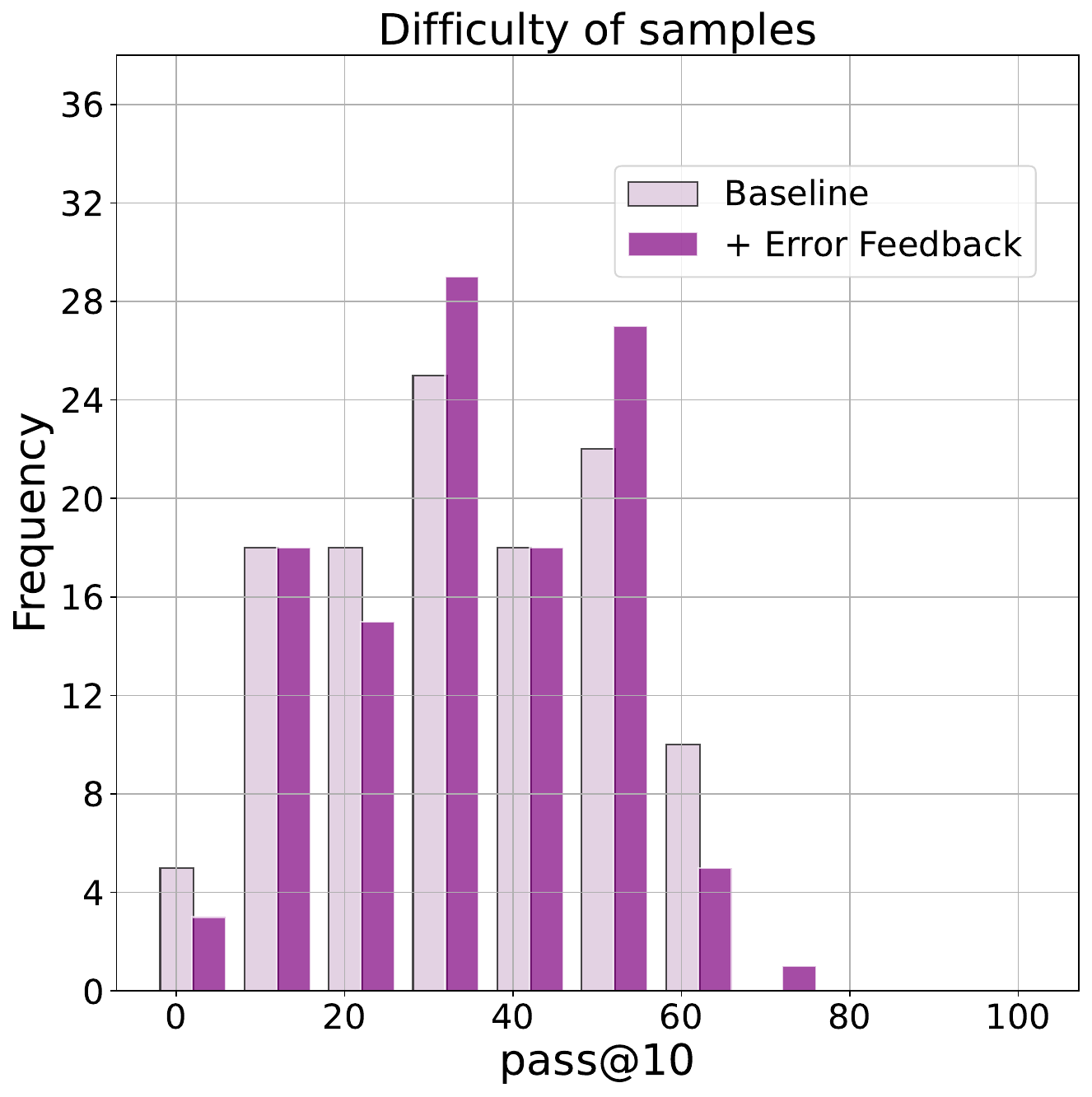}
        \label{fig:sample_diff_hist}
    \end{minipage}
    \hfill
    \begin{minipage}{0.49\linewidth}
        \centering
        \includegraphics[width=\linewidth]{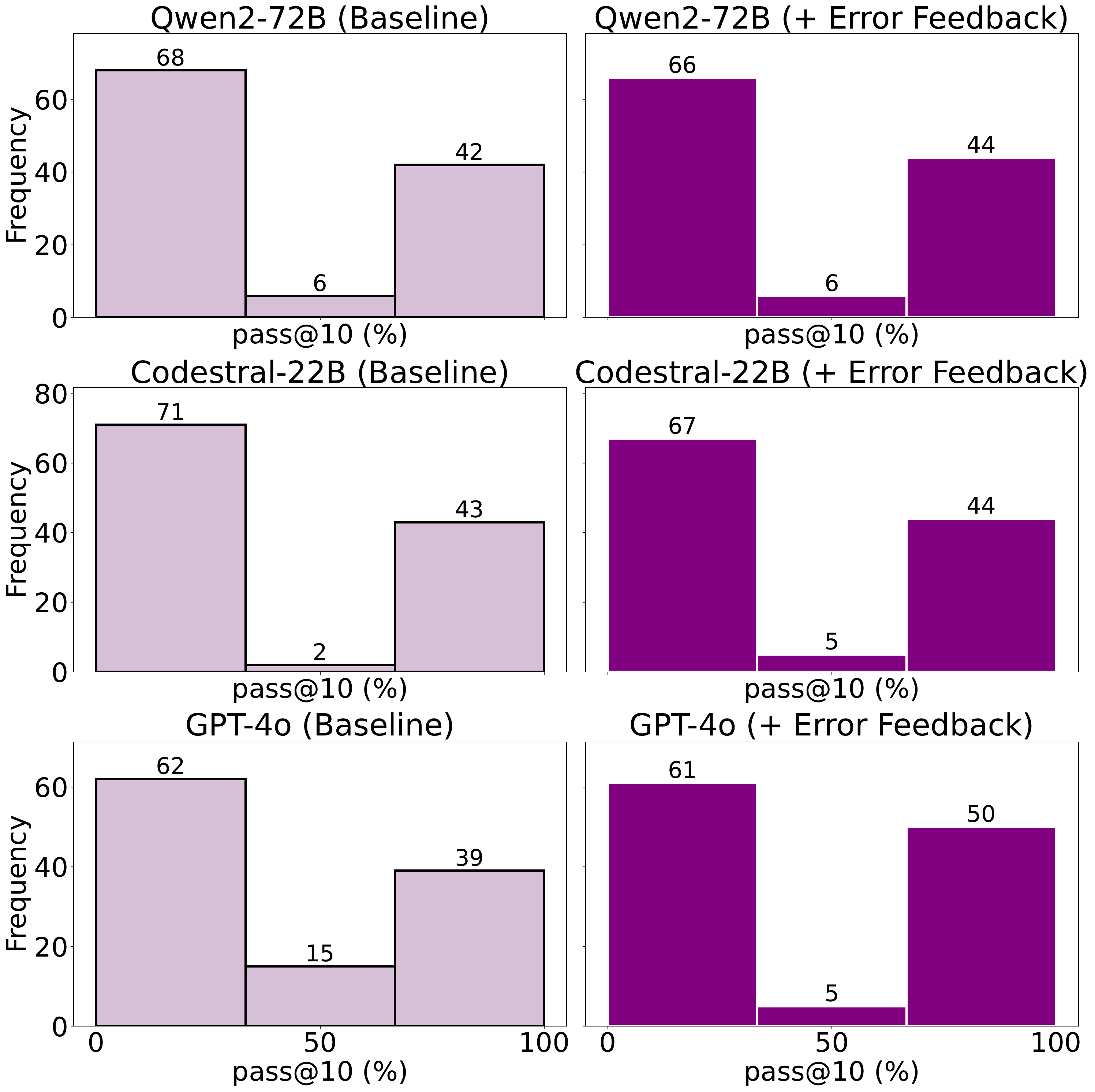}
        \label{fig:model_diff_hist}
    \end{minipage}
\caption{\textbf{Comparison of sample and model differences.} The left panel shows the distribution of sample difficulty, measured by the frequency of pass@10 scores across all models and seeds. The right panel presents the same distribution, but averaged for specific models across their seeds. Interestingly, individual models tend to exhibit bimodal distributions, indicating they are either consistently good or bad at specific problems. However, the aggregate distribution is less bimodal, suggesting that different models excel at different problems.}
\label{fig:combined_diff_hist}
\vspace{-3mm}
\end{figure}

Figure \ref{fig:combined_diff_hist} shows the distribution of sample difficulty. Notably, individual models (right panel) often display bimodal distributions, meaning they tend to perform consistently well or poorly on specific problems. In contrast, the aggregate distribution (left panel) is not bimodal, indicating that different models perform well on different sets of problems. The availability of error feedback shifts the distribution of the sample-wise difficulty to the right, as expected. Interestingly, some samples are not solved at all across models, even with feedback, and no samples are solved consistently by all models. As a further investigation, we plan to qualitatively examine samples that shift from unsolved to solved given error feedback. Finally, the right panel shows that many samples are either "easy" or "hard", however larger models tend to have more "medium" difficulty samples, indicating that scale can, at least partially, improve version-conditioned generation from unsolved to solved. We qualitatively demonstrate some of these examples in \ref{app:example_err_feedback}.

\begin{figure}[!ht]
    \centering
    \includegraphics[width=1.0\linewidth]{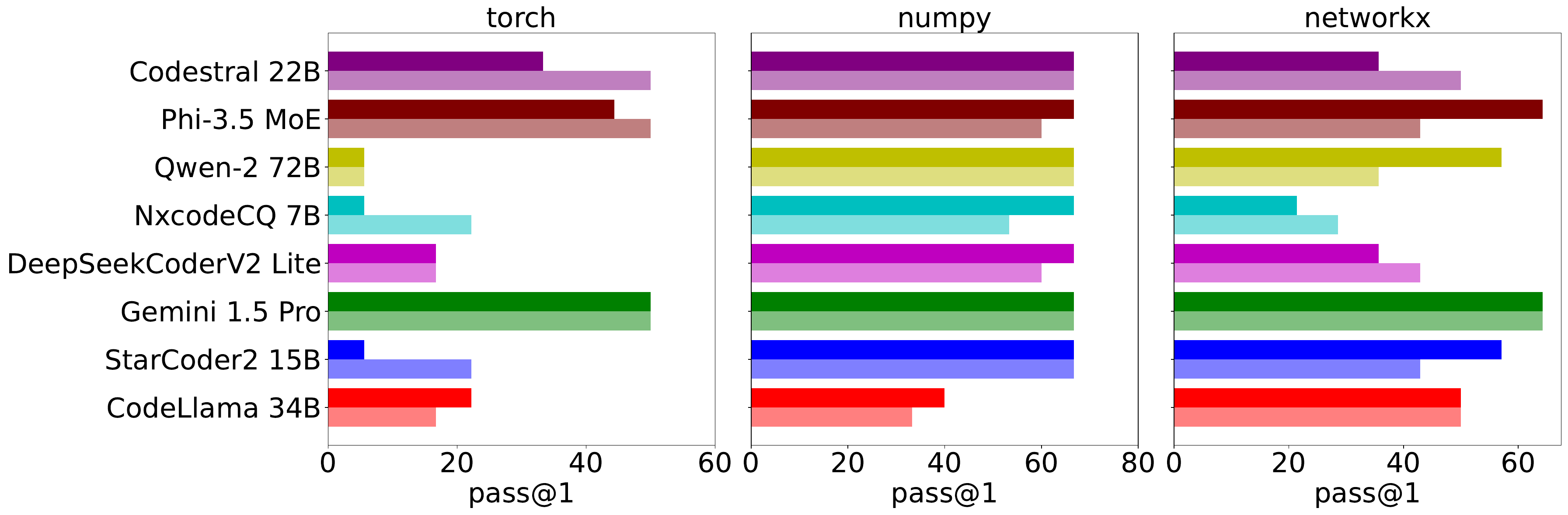}
    \caption{\textbf{Doc-prompting vs no doc (greedy decoding) instruct-tuned model performance on a subset of libraries.} Doc-prompting refers to the prompting technique in which the prompt contains the documentation of the library function specific to the problem at hand. Darker shaded bars represent doc-prompting values, lighter shaded bars represent no doc values.}
    \label{fig:oracle-greedy}
\end{figure}

\paragraph{Doc-prompting vs no doc} Figure \ref{fig:oracle-greedy} shows the instruct-tuned model performance breakdown between doc-prompting and no doc (standard prompt) with greedy decoding for a subset of libraries in the dataset. 
There is no consistent trend for any library, and in some cases, doc-prompting performs worse than the standard prompt (no doc). Moreover, performance across models is generally low for torch compared to the other libraries, and does not improve with doc-prompting. This suggests that the poor performance may be due to vague or difficult problem statements making it challenging for the model to interpret the problem.

\begin{wrapfigure}{O}{0.5\textwidth}
    \vspace{-5mm}
  \begin{center}
    \includegraphics[width=0.4\textwidth]{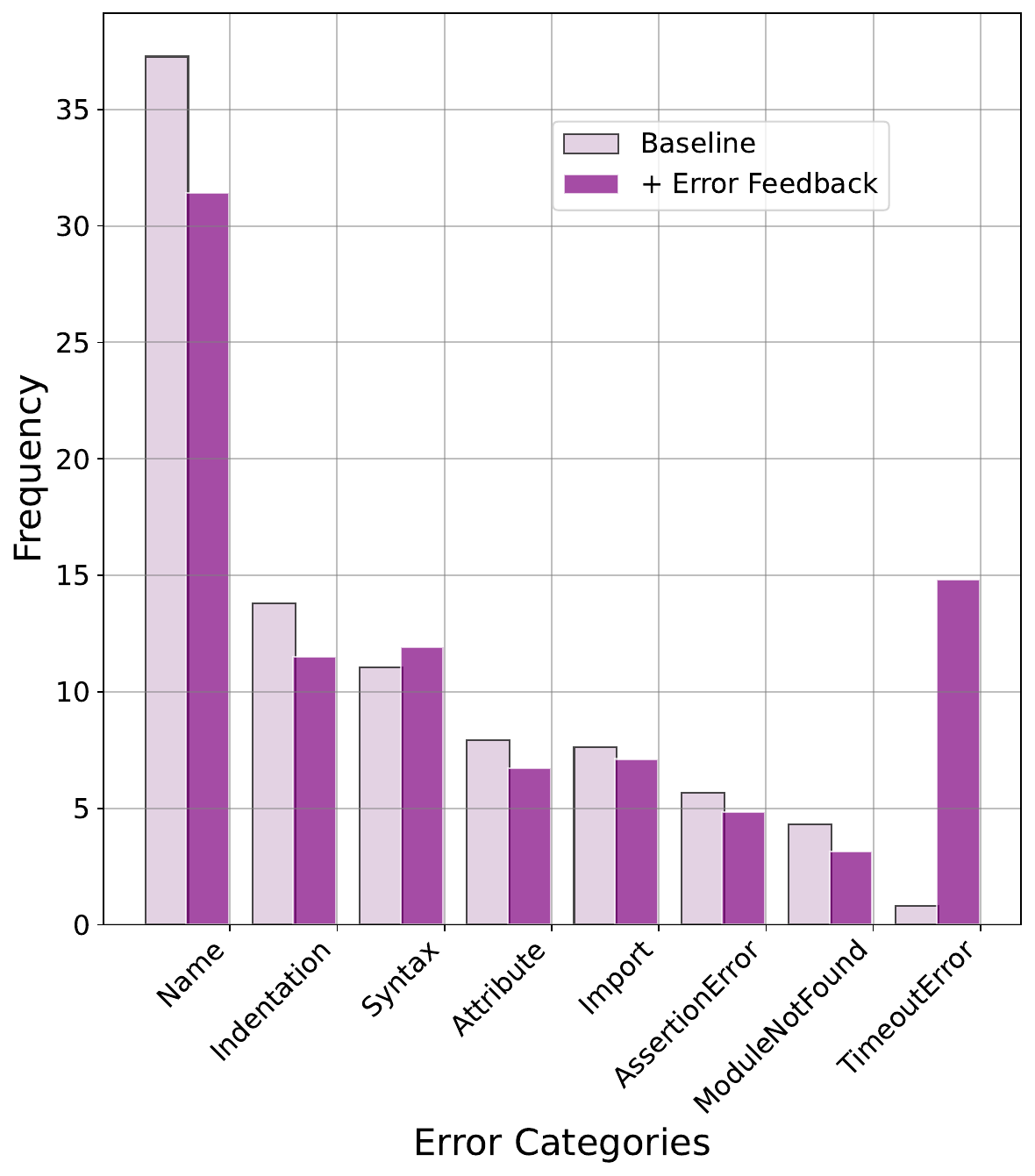}
  \end{center}
  \caption{\textbf{Most frequent categories of errors obtained from executing unit tests on model generated code, averaged across models.} The most common error by far is \textsc{Name} errors, which represents a variable that was used but not defined. This suggests that models do not grasp the context in many cases. Furthermore, \textsc{Name}, \textsc{Indentation}, \textsc{Attribute}, \textsc{Import}, \textsc{AssertionError}, and \textsc{ModuleNotFound} errors were all reduced with error feedback, while \textsc{TimeOut} and \textsc{Syntax} errors increased.}
  \label{fig:error-trace}
  \vspace{-10mm}
\end{wrapfigure}



\paragraph{Error Categorization} The objective of error feedback prompting is to allow models to self-correct by interpreting the error trace from the initial execution of the generated solution. Figure~\ref{fig:error-trace} reveals that across all instruction-based models, the most frequent error type is \textsc{Name}, indicating a gap in contextual awareness within these models. Although error feedback prompting reduces the occurrence of most types of errors, it unexpectedly increases the frequency of \textsc{Syntax} and \textsc{Timedout} errors.

\section{Related work}

\paragraph{Code LLM training and evaluation protocols}
\label{sec: codegen models}

Code LLM evaluations mainly revolve around code completion (\cite{zhang_repocoder_2023, vandam2023enriching, lu2021codexglue}). Existing benchmarks emphasize generic code completion, yet a recognized limitation is the inability of code LLM to generate and complete code that requires library and project-level knowledge \citep{xu_survey_2022}, let alone version-level knowledge, which is vital for real-world software applications.

Recent initiatives address repository-level code understanding by LLMs (\cite{bairi_codeplan_2023, shrivastava_repofusion_2023, shrivastava_repository-level_2023, liu_repobench_2023, deepseek-coder}). Attempts at library-level code generation (\cite{zan_cert_2022}) and consideration of dependencies between files (\cite{deepseek-coder}) have been made. However, these efforts do not directly address the challenge of accommodating version-sensitive changes, adding complexity.

The core issue arises from models being trained on library code without explicit knowledge of library versions or their functional changes. Consequently, when tasked with generating code specifically compatible with a particular library version, there is a significant risk models often encounter failures.

\paragraph{Datasets}

Existing datasets like HumanEval \citep{chen2021codex}, MBPP \citep{austin2021program}, and MTPB \citep{Nijkamp2022ACP} provide sets of handwritten prompts and test cases to evaluate code generated by code LLM. However, these datasets are relatively small and lack context regarding a model's comprehension of repositories. APPS \citep{hendrycks2021measuring} and CodeContest \citep{Li_2022} offer challenging datasets with coding competition questions, providing insights into a model's performance on difficult problems but without a focus on library-specific challenges. DSP \citep{chandel2022training} and DS-1000 \citep{lai2023ds} concentrate on the top data science libraries in Python, while JulCe \citep{agashe2019juice} uses Jupyter Notebooks for training and evaluation, but these notebooks do not necessarily need to be repository-specific. CoNaLa \citep{yin2018learning} contains problems collected from StackOverflow across multiple programming languages, including both library-specific questions and non-library-specific code. More recently, BigCodeBench~\citep{zhuo2024bigcodebench} is constructed to evaluate the comprehensive capabilities of code generation with tool use and instruction following, which poses a great challenge for existing models. Several datasets include version-specific code, such as Versicode~\citep{wu2024versicodeversioncontrollablecodegeneration}, CodeUpdateArena~\citep{liu2024codeupdatearenabenchmarkingknowledgeediting}, and the dataset by Wang et al. Versicode’s dataset, compiled from academic papers, Stack Overflow, and library source code, supports tasks like token, line, and block-level code completion and code editing. Unlike our dataset, Versicode evaluates using exact matches. Wang et al.’s dataset collects API mappings, such as “torch.lstsq() is deprecated in favor of torch.linalg.lstsq(),” and evaluates LLMs using exact match, edit similarity, and fixed rate metrics. Although Versicode and Wang et al.’s datasets address the evolving nature of libraries, their evaluations are limited to string matching.

In contrast, CodeUpdateArena evaluates LLMs’ ability to adapt to API changes, such as adding a boolean flag, by running tests. However, the dataset is synthetic and are not extracted from real-life version changes. For CodeUpdateArena, they also take the approach of training LLMs using the updated API function --using docstrings or examples--. It then tests if without access to the update during inference, the LLM's reflects the synthetic changes. While these datasets provide valuable resources for training and evaluating models, our \GitChameleon{} dataset advances research into version-conditioned code generation by LLMs. Runnable tests offer insights into LLM adaptability, and \GitChameleon{} further assesses a model’s ability to differentiate between library versions, and successfully use a specific version.
\paragraph{Implications for Lifelong Learning}

Continual/lifelong learning in code generation models is in its early stages \citep{yadav_exploring_2023,weyssow2023usage,wu2024continual,gao2023keeping}. However, current efforts often focus on artificial sequential tasks rather than utilizing the natural distribution shift in the chronological evolution of code. Notably, continual learning mainly targets mitigating catastrophic forgetting and balancing forward- and backward-transfer on a data stream, which may not align optimally with coding environment demands. In coding environments, obsolete or legacy libraries may prompt selective forgetting of irrelevant knowledge, particularly at the library/package level. Previous work, like \citet{caccia_online_2021}, which explores continual adaptation to shifting data distributions, and \citet{bornschein2022nevis}, which focuses on rapid adaptation to new distributions, could serve as foundations for enhancing continual learning in LLMs for code. We anticipate that \GitChameleon{} will inspire advancements in LLMs’ continual version adaptability, guiding development in this area. Such adaptability would be highly valuable in practical applications like enterprise-level code migration.

\section{Limitations}
We consider the lack of prompt optimization done for the instruct models as a considerable limitation of our analysis. 
Furthermore, the dataset consists of 116 problems, which is relatively small compared to other code benchmarks. 
Finally, we do not explore approaches such as RAG, chain of thought, or finetuning on a split of our benchmark to observe an upper bound of performance on this task. Future work could explore such approaches using our dataset.

\section{Conclusion}
Recognizing the crucial need for code LLM adaptation to evolving code environments, particularly in widely used libraries, we introduce a novel and extensive Python-based version-specific benchmark named \GitChameleon{}. 
By effectively leveraging \GitChameleon{}, we expose the shortcomings of existing state-of-the-art (SoTA) models in producing version-specific code, representing an inaugural effort to draw attention to this challenge. While our work exposes this shortcoming, we acknowledge the dataset's limitations. In future endeavors, we aim to enhance the dataset's comprehensiveness across various programming languages and frameworks. Additionally, we plan to introduce new tasks that can benefit research on code LLM models using \GitChameleon{}.

\section*{Acknowledgments}
The authors would like to express their sincere appreciation to Alex Gu for his insightful feedback, which significantly contributed to the development of the concept of the project. Furthermore, we extend our gratitude to Victor May for his technical assistance and expertise in supporting the benchmarking process. The authors acknowledge Bharat Runwal for his assistance during the data curation process and;  Deepanway Ghosal and Antonio Orvieto for their valuable feedback on the manuscript. This research was enabled in part by compute resources provided by Mila (\url{mila.quebec}) and the Max Planck Institut für Intelligente Systeme (\url{is.mpg.de}). The authors acknowledge support from the Canada CIFAR AI Chair Program and from the Canada Excellence Research Chairs (CERC) Program. 

\clearpage

\bibliography{arxiv}
\bibliographystyle{iclr2025_conference}

\clearpage

\appendix
\section{Appendix}

\subsection{Example of Doc-prompting}
\begin{lstlisting}[language=Python, caption=Sample pytorch problem with documentation]
You are to solve this in python using torch-1.11.0. Provide a self-contained Python script that solves the following problem in a markdown code block.
Calculate the natural logarithm of the absolute value of the gamma function using PyTorch`s special functions if available in this version, otherwise you may use another library.
Here is the documentation for the function to be used:
```
gammaln(x, /, out=None, *, where=True, casting='same_kind', order='K', dtype=None, subok=True[, signature, extobj])

gammaln(x, out=None)

Logarithm of the absolute value of the gamma function.

Defined as

.. math::

   \ln(\lvert\Gamma(x)\rvert)

where :math:`\Gamma` is the gamma function. For more details on
the gamma function, see [dlmf]_.

Parameters
----------
x : array_like
    Real argument
out : ndarray, optional
    Optional output array for the function results

Returns
-------
scalar or ndarray
    Values of the log of the absolute value of gamma

See Also
--------
gammasgn : sign of the gamma function
loggamma : principal branch of the logarithm of the gamma function

Notes
-----
It is the same function as the Python standard library function
:func:`math.lgamma`.

When used in conjunction with `gammasgn`, this function is useful
for working in logspace on the real axis without having to deal
with complex numbers via the relation ``exp(gammaln(x)) =
gammasgn(x) * gamma(x)``.

For complex-valued log-gamma, use `loggamma` instead of `gammaln`.

References
----------
.. [dlmf] NIST Digital Library of Mathematical Functions
          https://dlmf.nist.gov/5

Examples
--------
>>> import numpy as np
>>> import scipy.special as sc

It has two positive zeros.

>>> sc.gammaln([1, 2])
array([0., 0.])

It has poles at nonpositive integers.

>>> sc.gammaln([0, -1, -2, -3, -4])
array([inf, inf, inf, inf, inf])

It asymptotically approaches ``x * log(x)`` (Stirling`s formula).

>>> x = np.array([1e10, 1e20, 1e40, 1e80])
>>> sc.gammaln(x)
array([2.20258509e+11, 4.50517019e+21, 9.11034037e+41, 1.83206807e+82])
>>> x * np.log(x)
array([2.30258509e+11, 4.60517019e+21, 9.21034037e+41, 1.84206807e+82])
```

Please start with the following markdown code block:
```
import torch
input_tensor = torch.linspace(0, 10, steps=10)
# put answer in variable called output

```
    
\end{lstlisting}

\subsection{Qualitative Samples of Error Feedback}
\label{app:example_err_feedback}

\begin{lstlisting}[language=Python, caption=Original output from Codestral-22B-v0.1]
# [problem]
Calculate the natural logarithm of the absolute value of the gamma function using pytorch special functions if available in this version, otherwise you may use another library.
# [starting code]
import torch
input_tensor = torch.linspace(0, 10, steps=10)
output = 
# [sample solution]
torch.special.gammaln(input_tensor)

# [model output]
import torch
input_tensor = torch.linspace(0, 10, steps=10)
output = torch.log(torch.abs(torch.special.gamma(input_tensor)))
# [Error trace]
Traceback (most recent call last):
    output = torch.log(torch.abs(torch.special.gamma(input_tensor)))
AttributeError: module 'torch.special' has no attribute 'gamma'. Did you mean: 'digamma'?

\end{lstlisting}

\begin{lstlisting}[language=Python, caption=Regenerated solution with Codestral-22B-v0.1 given error feedback]
import torch
def natural_log_abs_gamma(input_tensor):
    gamma_tensor = torch.lgamma(input_tensor)
    natural_log_abs_gamma_tensor = torch.abs(gamma_tensor)
    return natural_log_abs_gamma_tensor
input_tensor = torch.linspace(0, 10, steps=10)
output = natural_log_abs_gamma(input_tensor)
\end{lstlisting}

\begin{lstlisting}[language=Python, caption=Original output from Phi-3.5-MoE-instruct]
# torch-1.10.0
# [problem]
Calculate the error function using pytorch special functions if available in this version, otherwise you may use another library.
# [starting code]
import torch\ninput_tensor = torch.linspace(0, 10, steps=10)\noutput = 
# [sample solution]
torch.special.erf(input_tensor)

# [model output]
import torch
from scipy.special import erf
input_tensor = torch.linspace(0, 10, steps=10)
def calculate_erf(tensor):
    # Convert tensor to numpy array
    numpy_array = tensor.numpy()
    # Calculate erf using scipy.special.erf
    output = erf(numpy_array)
    # Convert numpy array back to tensor
    output_tensor = torch.from_numpy(output)
    return output_tensor
output = calculate_erf(input_tensor)
# [Error trace]
Traceback (most recent call last):
    from scipy.special import erf
ModuleNotFoundError: No module named 'scipy'

\end{lstlisting}

\begin{lstlisting}[language=Python, caption=Regenerated solution with Phi-3.5-MoE-instruct given error feedback]
import torch
def compute_error_function(input_tensor):
    output = torch.erf(input_tensor)
    return output
input_tensor = torch.linspace(0, 10, steps=10)
output = compute_error_function(input_tensor)
\end{lstlisting}

\begin{lstlisting}[language=Python, caption=Original output with Phi-3.5-MoE-instruct]
# numpy-1.21.0
# [problem]
Write a function that rounds an array of numbers.
# [starting code]
import numpy as np

def custom_round(arr):
    return 
# [sample solution]
np.round_(arr)
# [model output]

import numpy as np
def custom_round(arr):
    return np.round(arr)
rounded_arr = custom_round(arr)


def test_custom_round():
    arr = np.array([1.5, 2.3, 3.7])
    result = custom_round(arr)
    expected = np.round_(arr)
    assert np.array_equal(result, expected)

test_custom_round()
# [Error trace]
Traceback (most recent call last):
   rounded_arr = custom_round(arr)
NameError: name 'arr' is not defined

\end{lstlisting}

\begin{lstlisting}[language=Python, caption=Regenerated solution with Phi-3.5-MoE-instruct given error feedback]
import numpy as np
def custom_round(arr):
    return np.round(arr)
\end{lstlisting}

\begin{lstlisting}[language=Python, caption=Original output with Phi-3.5-MoE-instruct]
# numpy-1.25.0
# [problem]
Write a function that rounds an array of numbers.
# [starting code]
import numpy as np

def custom_round(arr):
    return 
# [sample solution]
np.round(arr)
# [model output]

import numpy as np
def custom_round(arr):
    return np.round_(arr)


def test_custom_round():
    arr = np.array([1.5, 2.3, 3.7])
    result = custom_round(arr)
    expected = np.round(arr)
    assert np.array_equal(result, expected)

test_custom_round()
# [Error trace]
DeprecationWarning: `round_` is deprecated as of NumPy 1.25.0, and will be removed in NumPy 2.0. Please use `round` instead.
  result = custom_round(arr)

\end{lstlisting}

\begin{lstlisting}[language=Python, caption=Regenerated solution with Phi-3.5-MoE-instruct given error feedback]
import numpy as np
def custom_round(arr):
    return np.round(arr)
\end{lstlisting}

\begin{lstlisting}[language=Python, caption=Original output with Phi-3.5-MoE-instruct]
# pandas-1.4.0
# [problem]
Predict the correct type.
# [starting code]
import pandas as pd\nindex = pd.Index([1, 2, 3], dtype='int32')\nis_correct_type = index.dtype ==
# [sample solution]
 'int64'
# [model output]

import pandas as pd
def check_dtype(index):
is_correct_type = check_dtype(index)
# [Error trace]
Sorry: IndentationError: expected an indented block after function definition on line 3
\end{lstlisting}

\begin{lstlisting}[language=Python, caption=Regenerated solution with Phi-3.5-MoE-instruct given error feedback]
import pandas as pd
def is_correct_type(index, expected_dtype):
    return index.dtype == expected_dtype
index = pd.Index([1, 2, 3], dtype='int32')
\end{lstlisting} 

\begin{lstlisting}[language=Python, caption=Original output with Phi-3.5-MoE-instruct]
# gradio-2.9.2
# [problem]
Write a function that takes an image input and returns a textbox output.
# [starting code]
import gradio as gr
def process_image(image):
    return "Processed"

iface = gr.Interface
# [sample solution]
(fn=process_image, inputs=gr.inputs.Image(), outputs=gr.outputs.Textbox())
# [model output]

import gradio as gr
def process_image(image):
    # Perform processing on the image here
    # For simplicity, we're just returning the image dimensions
    width, height = image.shape[1], image.shape[0]
    return f"Width: {width}, Height: {height}"
iface = gr.Interface(
    fn=process_image,
    inputs=gr.Image(),
    outputs="text"
)
# [Error trace]
Traceback (most recent call last):
    inputs=gr.Image(),
AttributeError: module 'gradio' has no attribute 'Image'

\end{lstlisting}

\begin{lstlisting}[language=Python, caption=Regenerated solution with Phi-3.5-MoE-instruct given error feedback]
import gradio as gr
from PIL import Image
import numpy as np
def process_image(image):
    # Convert gradio image to PIL Image
    pil_image = Image.open(image)

    # Process the image (example: convert to grayscale)
    processed_image = pil_image.convert('L')

    # Convert processed image back to numpy array
    processed_image_array = np.array(processed_image)

    # Return the processed image array as a string
    return str(processed_image_array.tolist())
iface = gr.Interface(
    fn=process_image,
    inputs=gr.inputs.Image(shape=(None, None, 3)),
    outputs=gr.outputs.Textbox()
)
\end{lstlisting}

\begin{lstlisting}[language=Python, caption=Original output with Phi-3.5-MoE-instruct]
# networkx-2.5
# [problem]

Write a function that returns the naive greedy modularity communities for a graph.

# [starting code]
import networkx as nx
def naive_modularity_communities(G):
    return nx.community.
# [sample solution]
naive_greedy_modularity_communities(G)
# [model output]

import networkx as nx
def naive_modularity_communities(G):
    communities = []
    while G.number_of_nodes() > 0:
        node = max(G.degree, key=lambda x: x[1])[0]
        community = nx.algorithms.community.greedy_modularity_communities(G, nbunch=[node])
        communities.append(community)
        G.remove_nodes_from(community)
    return communities
G = nx.karate_club_graph()
communities = naive_modularity_communities(G)

G = nx.karate_club_graph()
# [Error trace]
Traceback (most recent call last):
    community = nx.algorithms.community.greedy_modularity_communities(G, nbunch=[node])
TypeError: greedy_modularity_communities() got an unexpected keyword argument 'nbunch'

\end{lstlisting}

\begin{lstlisting}[language=Python, caption=Regenerated solution with Phi-3.5-MoE-instruct given error feedback]
import networkx as nx
def naive_modularity_communities(G):
    communities = []
    unvisited_nodes = list(G.nodes)
    
    while unvisited_nodes:
        current_node = unvisited_nodes[0]
        community = nx.algorithms.community.greedy_modularity_communities(G, start_nodes=[current_node])
        visited_nodes = list(community[0])
        unvisited_nodes = list(set(unvisited_nodes) - set(visited_nodes))
        communities.append(visited_nodes)
    
    return communities
G = nx.karate_club_graph()
result = naive_modularity_communities(G)
\end{lstlisting}

\begin{lstlisting}[language=Python, caption=Original output with gpt4o]
# torch-1.10.0
# [problem]
Calculate the complementary error function using pytorch's special functions if available in this version, otherwise you may use another library.
# [starting code]
import torch\ninput_tensor = torch.linspace(0, 10, steps=10)\noutput = 
# [sample solution]
torch.special.erfc(input_tensor)
# [model output]

import torch
import scipy.special
def complementary_error_function(x):
    # Check if 'special.erfc' is available in the current version of PyTorch
    if hasattr(torch.special, 'erfc'):
        return torch.special.erfc(x)
    else:
        # Use scipy's erfc function if PyTorch's is not available
        x_np = x.detach().numpy()  # Convert the tensor to a numpy array
        erfc_np = scipy.special.erfc(x_np)  # Calculate erfc using scipy
        return torch.from_numpy(erfc_np)  # Convert the result back to a tensor
input_tensor = torch.linspace(0, 10, steps=10)
output = complementary_error_function(input_tensor)
# [Error trace]
Traceback (most recent call last):
    import scipy.special
ModuleNotFoundError: No module named 'scipy'

\end{lstlisting}

\begin{lstlisting}[language=Python, caption=Regenerated solution with gpt4o given error feedback]
import torch
def complementary_error_function(x):
    # Check if 'special.erfc' is available in the current version of PyTorch
    if hasattr(torch.special, 'erfc'):
        return torch.special.erfc(x)
    else:
        # Raise an error if PyTorch's special.erfc is not available and scipy is not installed
        raise ImportError("PyTorch version does not support special.erfc and scipy is not installed.")
input_tensor = torch.linspace(0, 10, steps=10)
output = complementary_error_function(input_tensor)
\end{lstlisting}

\begin{lstlisting}[language=Python, caption=Original output with gpt4o]
# scikit-learn-1.3
# [problem]
Retrieve and list all available scorer names, ensuring they are returned in a list format.
# [starting code]
from sklearn import metrics\nscorer_names = 
# [sample solution]
metrics.get_scorer_names()
# [model output]

from sklearn import metrics
scorer_names = list(metrics.SCORERS.keys())
# [Error trace]
Traceback (most recent call last):
AttributeError: module 'sklearn.metrics' has no attribute 'SCORERS'

\end{lstlisting}

\begin{lstlisting}[language=Python, caption=Regenerated solution with gpt4o given error feedback]
from sklearn.metrics import get_scorer_names
scorer_names = list(get_scorer_names())
conditions = isinstance(scorer_names, list) and len(scorer_names) > 0
\end{lstlisting}
\begin{lstlisting}[language=Python, caption=Original output with gpt4o]
# gradio-3.36.0
# [problem]
Write a function that displays an image using Gradio where you cannot share the image.
# [starting code]
import gradio as gr
def display_image():
    return "https://image_placeholder.com/42"

iface = gr.Interface
# [sample solution]
(fn=display_image, inputs=[], outputs=gr.Image(show_share_button=False))

# [model output]

import gradio as gr
def display_image():
    # This is just a placeholder URL; replace it with your actual image URL or file path.
    return "https://via.placeholder.com/150"
iface = gr.Interface(
    fn=display_image,
    inputs=None,
    outputs=gr.Image(label="Display Image", type="auto"),
    live=False  # Ensure live updates are off
)
# [Error trace]
 line 116, in __init__
    raise ValueError(
ValueError: Invalid value for parameter `type`: auto. Please choose from one of: ['numpy', 'pil', 'filepath']

\end{lstlisting}

\begin{lstlisting}[language=Python, caption=Regenerated solution with gpt4o given error feedback]
import gradio as gr
def display_image():
    # This is just a placeholder URL; replace it with your actual image URL or file path.
    return "https://via.placeholder.com/150"
iface = gr.Interface(
    fn=display_image,
    inputs=None,
    outputs=gr.Image(label="Display Image", type="filepath"),
    live=False  # Ensure live updates are off
)
iface.share = False
\end{lstlisting}

\begin{lstlisting}[language=Python, caption=Original output with gpt4o]
# gradio-3.24.0
# [problem]
Write a function that takes an image input and returns a label output.
# [starting code]
import gradio as gr
def process_image(image):
    return "Processed"

iface = gr.Interface
# [sample solution]
(fn=process_image, inputs=gr.Image(), outputs=gr.Label())
# [model output]

import gradio as gr
def process_image(image):
    # Placeholder function to process the image and return a label
    # You can add your image processing and model inference logic here
    return "Processed"
iface = gr.Interface(
    fn=process_image,         # The function to process the image
    inputs=gr.inputs.Image(), # The input component (an image)
    outputs=gr.outputs.Text() # The output component (a text label)
)
# [Error trace]
python3.10/site-packages/gradio/inputs.py:257: UserWarning: Usage of gradio.inputs is deprecated, and will not be supported in the future, please import your component from gradio.components
  warnings.warn(
python3.10/site-packages/gradio/deprecation.py:40: UserWarning: `optional` parameter is deprecated, and it has no effect
  warnings.warn(value)
Traceback (most recent call last):
    outputs=gr.outputs.Text() # The output component (a text label)
AttributeError: module 'gradio.outputs' has no attribute 'Text'

\end{lstlisting}

\begin{lstlisting}[language=Python, caption=Regenerated solution with gpt4o given error feedback]
import gradio as gr
def process_image(image):
    # Placeholder function to process the image and return a label
    # You can add your image processing and model inference logic here
    return "Processed"
iface = gr.Interface(
    fn=process_image,          # The function to process the image
    inputs=gr.Image(),         # The input component (an image)
    outputs=gr.Label()         # The output component (a text label)
)
\end{lstlisting}

\begin{lstlisting}[language=Python, caption=Original output with gpt4o]
# networkx-2.5
# [problem]

Write a function that accesses the first edge of a NetworkX graph.

# [starting code]
import networkx as nx
def get_first_edge(G):
    return 
# [sample solution]
list(G.edges)[0]
# [model output]

import networkx as nx
def get_first_edge(G):
    # Check if the graph has any edges
    if G.number_of_edges() == 0:
        return None
    # Access the first edge
    first_edge = next(G.edges(data=True))
    return first_edge

G = nx.karate_club_graph()
# [Error trace]
Traceback (most recent call last):
    line 8, in get_first_edge
    first_edge = next(G.edges(data=True))
TypeError: 'EdgeDataView' object is not an iterator

\end{lstlisting}

\begin{lstlisting}[language=Python, caption=Regenerated solution with gpt4o given error feedback]
import networkx as nx
def get_first_edge(G):
    # Check if the graph has any edges
    if G.number_of_edges() == 0:
        return None
    # Access the first edge
    first_edge = list(G.edges(data=True))[0]
    return first_edge
G = nx.karate_club_graph()
first_edge = get_first_edge(G)
\end{lstlisting}

\subsection{Pseudocode for our verificaiton process}
\label{app:verification_pseudocode}

\begin{lstlisting}[language=Python]
# Create and activate virtual environment
Run: "python -m venv venv"
Run: "source venv/bin/activate"

# Install specified library and version
Run: "pip install $library==$version"

# Install additional dependencies if specified
If additional_dep:
    Run: "pip install $additional_dep"

# Combine code snippets
complete_code = starter_code + expected_output + test

# Run the combined code with a timeout
Run: "timeout 60 python -c '$complete_code'"

# Capture and print exit code
exit_code = LastCommandExitCode()
Print: "THIS WAS THE EXIT CODE: $exit_code"

# Print the complete code
Print: complete_code

# Deactivate and remove virtual environment
Run: "deactivate"
Run: "rm -rf venv"
\end{lstlisting}

Each sample was validated using this method to ensure that it functioned as intended.

\end{document}